\documentclass[%
groupedaddress,
 amsmath,amssymb,
]{revtex4-2}

\usepackage{graphicx}
\usepackage{dcolumn}
\usepackage{bm}
\usepackage{hyperref}

\usepackage{physics}

\usepackage{academicons}
\usepackage[dvipsnames]{xcolor}
\newcommand{\orcid}[1]{\href{https://orcid.org/#1}{\includegraphics[height=3mm]{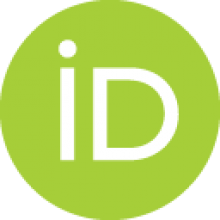}}}

\begin{document}

\title{Commissioning of a compact multibend achromat lattice: A new 3 GeV synchrotron radiation facility}

\newcommand{\qst}{\affiliation{National Institutes for Quantum Science and Technology (QST), Sendai, Miyagi 980-8572, Japan}}
\newcommand{\natco}{\affiliation{NAT Corporation, Naka, Ibaraki 319-1112, Japan}}
\newcommand{\ses}{\affiliation{SPring-8 Service Co., Ltd. (SES), Tatsuno, Hyogo 679-5165, Japan}}
\newcommand{\rikensp}{\affiliation{RIKEN SPring-8 Center (RSC), Sayo, Hyogo 679-5148, Japan}}
\newcommand{\jasri}{\affiliation{Japan Synchrotron Radiation Research Institute (JASRI), Sayo, Hyogo 679-5198, Japan}}
\newcommand{\kek}{\affiliation{KEK, Tsukuba, Ibaraki 305-0801, Japan}}


\author{Shuhei Obara~\orcid{0000-0003-3488-3553}} \email{obara.shuhei@qst.go.jp} \qst
\author{Kota Ueshima} \qst
\author{Takao Asaka} \qst
\author{Yuji Hosaka~\orcid{0000-0003-3084-6528}} \qst
\author{Koichi Kan} \qst
\author{Nobuyuki Nishimori} \qst

\author{Toshitaka Aoki} \natco \qst
\author{Hiroyuki Asano} \natco \qst
\author{Koichi Haga} \natco \qst
\author{Yuto Iba} \natco \qst
\author{Akira Ihara} \natco \qst
\author{Katsumasa Ito} \natco \qst
\author{Taiki Iwashita} \natco \qst
\author{Masaya Kadowaki} \natco \qst
\author{Rento Kanahama} \natco \qst
\author{Hajime Kobayashi} \natco \qst
\author{Hideki Kobayashi} \natco \qst
\author{Hideo Nishihara} \natco \qst
\author{Masaaki Nishikawa} \natco \qst
\author{Haruhiko Oikawa} \natco \qst
\author{Ryota Saida} \natco \qst
\author{Keisuke Sakuraba} \natco \qst
\author{Kento Sugimoto} \natco \qst
\author{Masahiro Suzuki} \natco \qst
\author{Kouki Takahashi} \natco \qst
\author{Shunya Takahashi} \natco \qst
\author{Tatsuki Tanaka} \natco \qst
\author{Tsubasa Tsuchiyama} \natco \qst
\author{Risa Yoshioka} \natco \qst

\author{Tsuyoshi Aoki~\orcid{0009-0008-9167-2107}} \jasri \qst
\author{Hideki Dewa} \jasri \qst
\author{Takahiro Fujita} \jasri \qst
\author{Morihiro Kawase} \jasri \qst
\author{Akio Kiyomichi~\orcid{0009-0008-3493-5302}} \jasri \qst
\author{Takashi Hamano} \jasri \qst
\author{Mitsuhiro Masaki~\orcid{0000-0002-3497-3017}} \jasri \qst
\author{Takemasa Masuda} \jasri \qst
\author{Shinichi Matsubara} \jasri \qst
\author{Kensuke Okada} \jasri \qst
\author{Choji Saji} \jasri \qst
\author{Tsutomu Taniuchi~\orcid{0000-0001-9193-4965}} \jasri \qst
\author{Yukiko Taniuchi} \jasri \qst
\author{Yosuke Ueda} \jasri \qst
\author{Hiroshi Yamaguchi~\orcid{0000-0002-7787-8116}} \jasri \qst
\author{Kenichi Yanagida} \jasri \qst

\author{Kenji Fukami~\orcid{0000-0002-1207-3237}} \jasri \rikensp \qst
\author{Naoyasu Hosoda} \jasri \rikensp \qst
\author{Miho Ishii} \jasri \rikensp \qst
\author{Toshiro Itoga~\orcid{0000-0002-5887-3707}} \jasri \rikensp \qst
\author{Eito Iwai~\orcid{0000-0001-8691-4396}} \jasri \rikensp \qst
\author{Tamotsu Magome} \jasri \rikensp \qst
\author{Masaya Oishi} \jasri \rikensp \qst
\author{Takashi Ohshima~\orcid{0000-0002-0192-8690}} \jasri \rikensp \qst
\author{Chikara Kondo} \jasri \rikensp \qst
\author{Tatsuyuki Sakurai} \jasri \rikensp \qst
\author{Masazumi Shoji} \jasri \rikensp \qst
\author{Takashi Sugimoto~\orcid{0000-0002-1911-998X}} \jasri \rikensp \qst
\author{Shiro Takano~\orcid{0000-0001-8759-9690}} \jasri \rikensp \qst

\author{Kazuhiro Tamura} \jasri \rikensp \qst
\author{Takahiro Watanabe~\orcid{0000-0001-8896-9788}} \jasri \rikensp \qst

\author{Takato Tomai~\orcid{0009-0003-9964-0381}} \jasri

\author{Noriyoshi Azumi} \jasri \rikensp

\author{Takahiro Inagaki~\orcid{0000-0003-4534-7887}} \rikensp \jasri
\author{Hirokazu Maesaka~\orcid{0000-0002-6205-5571}} \rikensp \jasri
\author{Sunao Takahashi} \rikensp \jasri
\author{Takashi Tanaka} \rikensp \jasri

\author{Shinobu Inoue} \ses
\author{Hirosuke Kumazawa} \ses
\author{Kazuki Moriya} \ses
\author{Kohei Sakai} \ses
\author{Toshio Seno} \ses
\author{Hiroshi Sumitomo} \ses
\author{Ryoichi Takesako} \ses
\author{Shinichiro Tanaka} \ses
\author{Ryo Yamamoto} \ses
\author{Kazutoshi Yokomachi} \ses
\author{Masamichi Yoshioka} \ses

\author{Toru Hara} \rikensp
\author{Sakuo Matsui} \rikensp
\author{Toshihiko Hiraiwa~\orcid{0000-0003-4251-2947}} \rikensp 
\author{Hitoshi Tanaka~\orcid{0000-0003-1848-2338}} \rikensp \qst

\author{Hiroyasu Ego~\orcid{0000-0002-9661-823X}} \kek


\date{\today}

\begin{abstract}
NanoTerasu, a new $3$\,GeV synchrotron light source in Japan, began user operation in April 2024. 
It provides high-brilliance soft to tender X-rays and covers a wide spectral range from ultraviolet to tender X-rays. 
Its compact storage ring with a circumference of $349\,{\rm m}$ is based on a four-bend achromat lattice to provide two straight sections in each cell for insertion devices with a natural horizontal emittance of $1.14\,{\rm nm}\,{\rm rad}$, which is small enough for soft X-rays users. 
The NanoTerasu accelerator incorporates several innovative technologies, including a full-energy injector C-band linear accelerator with a length of $110\,{\rm m}$, an in-vacuum off-axis injection system, a four-bend achromat with B-Q combined bending magnets, and a TM020 mode accelerating cavity with built-in higher-order-mode dampers in the storage ring.  
This paper presents the accelerator machine commissioning over a half-year period and our model-consistent ring optics correction. 
The first user operation with a stored beam current of $160\,{\rm mA}$ is also reported.
We summarize the storage ring parameters obtained from the commissioning. 
This is helpful for estimating the effective optical properties of synchrotron radiation at NanoTerasu.
\end{abstract}


\maketitle

\section{Introduction} \label{sec:intro}

NanoTerasu is a new $3$\,GeV fourth-generation synchrotron light source constructed on a green field site in Sendai, Japan. 
The purpose of this facility is to provide high-brilliance soft and tender X-rays, and to cover a wide spectral range from ultraviolet to tender X-rays, using a compact storage ring with a circumference of $349\,{\rm m}$ as a complementary partner to SPring-8~\cite{spring8_kamitsubo_1998}, which mainly covers hard X-rays and is one of the world's most powerful synchrotron radiation facilities.
The target brilliance of NanoTerasu is $\order{10^{21}}\,{\rm photons/s/mm^2/mrad^2/0.1\%\,bandwidth}$ with a coherence ratio of roughly 10\% for photon energies of $1$--$3\,{\rm keV}$.

The designed natural horizontal emittance is $1.14\,{\rm nm}\,{\rm rad}$ with a nominal stored current of $400\,{\rm mA}$, and
a four-bend achromat lattice is used in the NanoTerasu compact storage ring to meet the electron beam specification.
The fourth-generation synchrotron light sources based on the multi-bend achromat lattice proposed in 1993~\cite{einfeld_design_1993} have been in service at MAX-IV since 2016~\cite{tavares_commissioning_2018}, ESRF-Extremely Brilliant Source since 2020~\cite{raimondi_commissioning_2021}, and Sirius since 2021~\cite{sirius_status_2023}, and many others are under construction or planned~\cite{apsu_tdr, thai_3gev, spring8-II}. 
NanoTerasu started user operation in April 2024 as the world's fourth fourth-generation light source.

\begin{figure}[htbp]
    \centering
    \includegraphics[width=\linewidth]{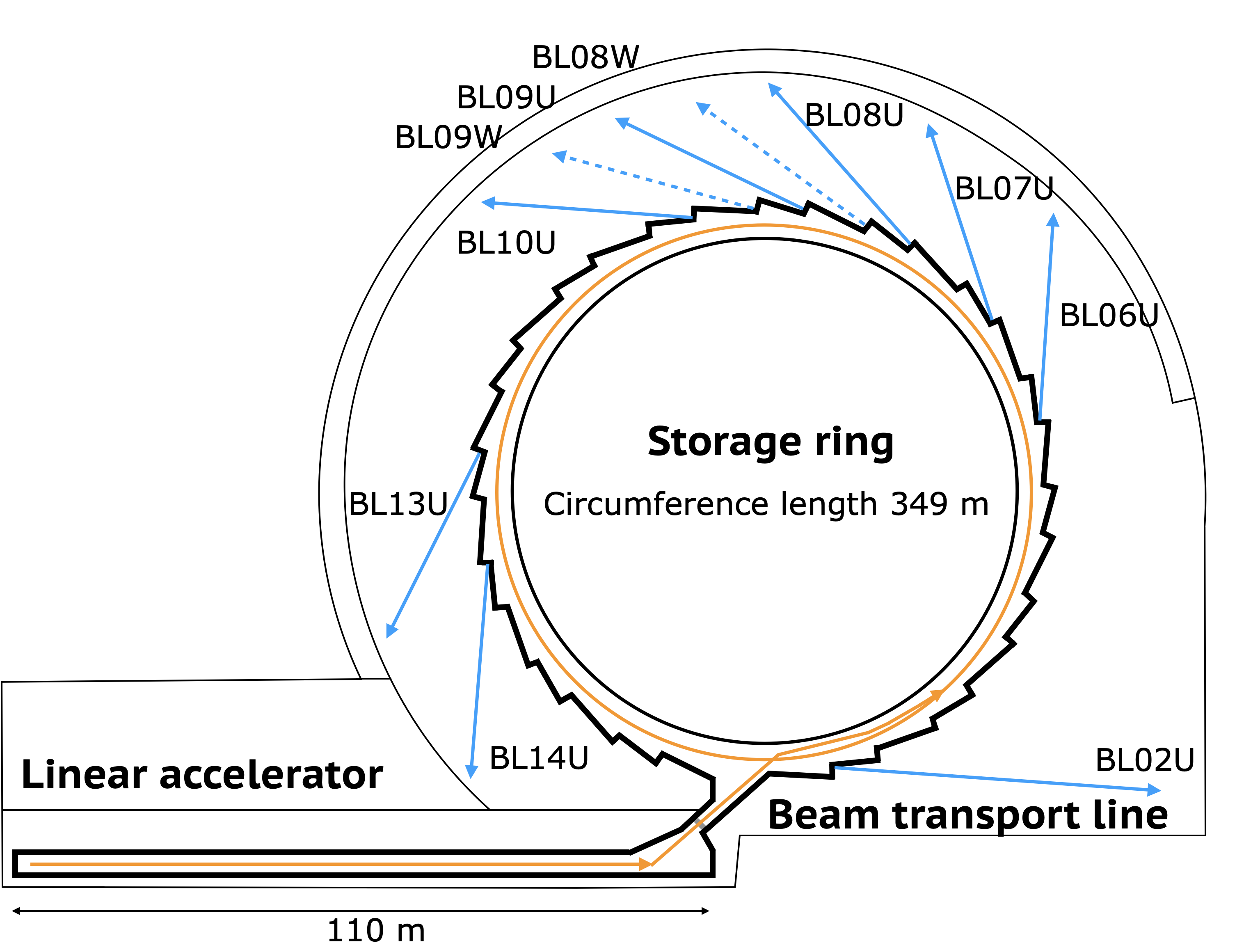}
    \caption{Schematics of NanoTerasu. 
    The 10 beamlines shown are the ones available in the first year. 
    Blue dashed lines represent MPW beamlines, and blue solid lines represent undulator beamlines. 
    The orange line shows the electron beam trajectory.}
    \label{fig:nanoterasu}
\end{figure}

A schematic of NanoTerasu is shown in Fig.~\ref{fig:nanoterasu}.
The light source can provide a maximum of 28 insertion device beamlines, although it will provide 10 beamlines in the initial user operation phase.
Two beamlines are multi-pole wiggler (MPW) beamlines, and the others consist of two in-vacuum, one twin-helical, and five APPLE-II undulator beamlines.
Various beamline designs have been proposed and are being constructed~\cite{nanoterasu_web,Miyawaki_2022,Horiba_2022,Ohtsubo_2022}. 
Moreover, the NanoTerasu facility has an upgrade plan to extend the linear accelerator beam lines and serve as a soft X-ray free-electron laser in the future~\cite{nanoterasu_cdr}.

In this paper, Sec.~\ref{sec:design} briefly describes the accelerator design, and Sec.~\ref{sec:commissioning} describes the efficient completion of the NanoTerasu accelerator machine commissioning in which model-consistent storage ring optics and a $200\,{\rm mA}$ stored beam were obtained within half a year.
We also report the first user service time operation in Sec.~\ref{sec:usertime}.

\section{Accelerator Design} \label{sec:design}
The NanoTerasu accelerator consists of a $3$\,GeV linear accelerator as an injector, a beam transport line, and a storage ring. 
The linear accelerator has a short length of $110\,{\rm m}$ because it uses a compact low-emittance radio-frequency (RF) electron gun system~\cite{asaka_low-emittance_2020} and C-band high accelerating gradient structures of more than $40\,{\rm MV/m}$.
The storage ring is also compact, with a circumference of $349\,{\rm m}$ compared with soft X-ray 3\,GeV synchrotron facilities worldwide~\cite{tavares_commissioning_2018,sirius_status_2023}.
To achieve a horizontal emittance of $1.14\,{\rm nm}\,{\rm rad}$ with the compact storage ring, the lattice design is a four-bend achromat with 16 cells; thus, there are 16 long straight sections and 16 short straight sections.
Undulators and MPWs are installed in the long and short straight sections, respectively.
Two long straight sections are used for beam injection and ring RF cavities, and two short straight sections are used for a stored beam current monitor and a bunch-by-bunch feedback system. 
This section briefly describes the key points of the accelerator design; the details are in the conceptual design report~\cite{nanoterasu_cdr}.

\subsection{Linear accelerator and beam transport line} \label{subsec:linac_design}

\begin{figure*}[htb]
    \centering
    \includegraphics[width=\linewidth]{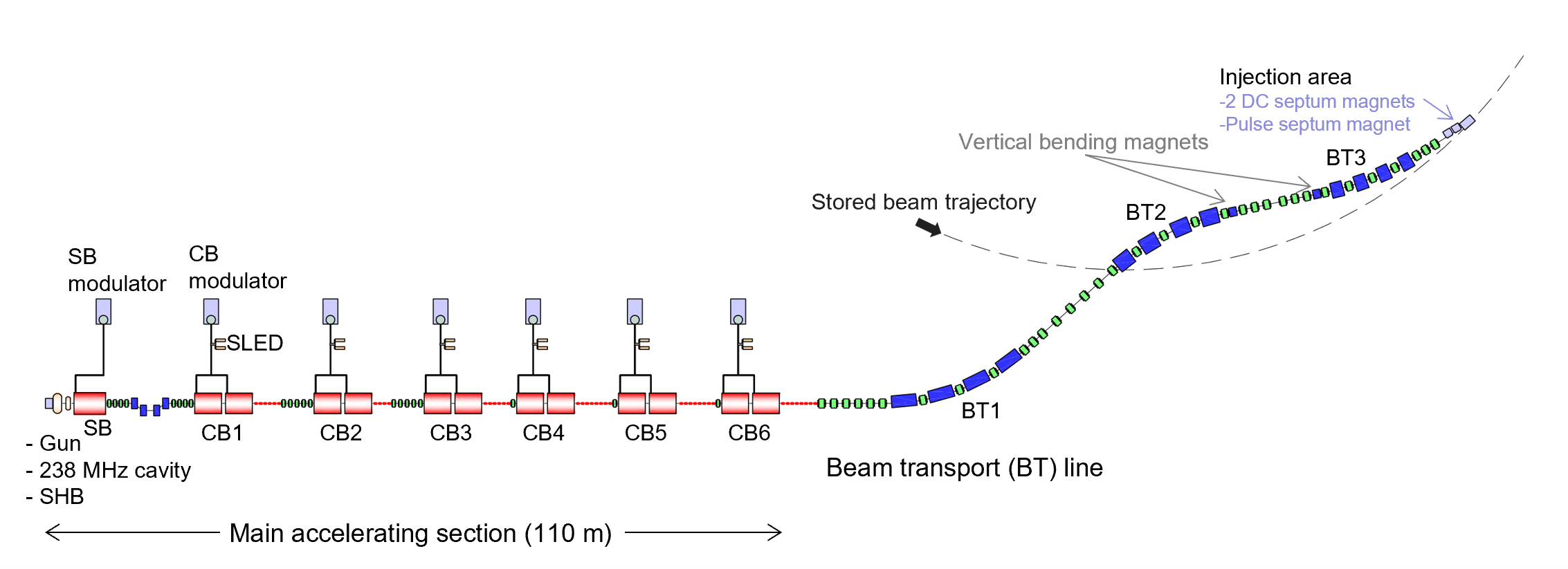}
    \caption{Diagram of the linear accelerator and beam transport line of NanoTerasu. 
    S-band (SB) and C-band (CB) modulators, and six CB sections with only each first unit are shown.
    The beam transport line consists of three sections from the exit of the linear accelerator to the injection area: BT1, BT2, and BT3. 
    Bending (blue) and quadrupole (green) magnets are shown.}
    \label{fig:linacBT}
\end{figure*}

Figure~\ref{fig:linacBT} shows a schematic of the NanoTerasu injector system, which consists of a 110-m-long $3\,{\rm GeV}$ linear accelerator section and an 80-m-long beam transport section. 
Stable beam injection into a low-emittance storage ring with a horizontal dynamic aperture of $-15\,{\rm mm}$ requires a low-emittance injector beam from the linear accelerator.

The emittance required for the 3\,GeV linear accelerator is less than $2\,{\rm nm}\,{\rm rad}$, corresponding to a normalized emittance of less than $10\,{\rm mm}\,{\rm mrad}$ with a bunch charge of $0.3\,{\rm nC}$~\cite{nanoterasu_cdr}.
A 50\,kV DC electron gun equipped with a gridded thermionic cathode (EIMAC Y845, CPI), which has been used in many accelerator facilities, was developed to satisfy the emittance requirements and user requirements such as equipment robustness and near-maintenance-free operation. 
Although the grid degrades emittance, the transparent grid condition achieved by the optimized cathode grid voltage allows the generation of a low-emittance beam~\cite{asaka_low-emittance_2020}.
Emittance growth due to space charge effects is suppressed by immediately increasing the electron beam energy from $50$ to $500\,{\rm keV}$ with a 238\,MHz accelerating RF cavity placed just downstream at the 50\,kV gun exit. 
The measured normalized emittance of the $500\,{\rm keV}$ electron beam is $1.7\,{\rm mm}\,{\rm mrad}$ in the core part which contains $60\%$ of the $1\,{\rm nC}$ bunch charge, satisfying the requirements as an electron source not only for the NanoTerasu injector but also for a soft X-ray free-electron laser.
The $500\,{\rm keV}$ electron beam is bunch-compressed by a 476\,MHz sub-harmonic buncher (SHB) from $500$ to $5\,{\rm ps}$ FWHM and accelerated up to $40\,{\rm MeV}$ in a 2-m-long S-band accelerating structure. 

The main acceleration section consists of 20 C-band accelerating units, which provide efficient acceleration from $40\,{\rm MeV}$ to $3\,{\rm GeV}$. 
In a single unit, the $50\,{\rm MW}$, $2.5\,\mu{\rm s}$ RF output from the klystron~\cite{inagaki_klystron_2014} is increased to $180\,{\rm MW}$ by an RF pulse compression cavity (SLED~\cite{farkas_sled_1974}) and fed into two 2-m-long C-band accelerating structures. 
The C-band accelerating units achieve an acceleration gradient of $42\,{\rm MV/m}$. 
A MicroTCA.4 system for precise low-level RF control of the C-band units is used to obtain high electron beam stability and reproducibility.  
The MicroTCA.4 system enables high-density implementation and high-speed signal processing.
It consists of a high-speed digitizer (250\,MS/s) and an RF front-end with 8-channel input and 1-channel vector modulated output.
The system monitors the RF power and phase of the low-level amplifier output, klystron input/output, SLED output, traveling wave output to the accelerating structure, and reflected wave. 
The RF frontend has adjustable RF attenuation from 0\,dB to 31.5\,dB in steps of 0.5\,dB, and compensation parameters for the phase difference due to this attenuator are measured in advance. 
The attenuator of the accelerating structure is set to 0\,dB during the beam-induced signal measurement shown in Sec.~\ref{subsec:linac_commissioning}.

The electron beam from the 3\,GeV linear accelerator is transported to the storage ring through an 80-m-long beam transport section. 
The electron beam is transported through three deflection sections horizontally and one deflection section vertically. 
The deflection sections are formed under an achromatic condition in which the dispersion function of each deflection section is locally closed on the order of several tens of millimeters. 
The beam transport line has a grade-separated crossing to the stored beam orbit with a different height and goes to the inside.
Two vertical bending magnets lift up the electron beam by a height of $600\,{\rm mm}$ between the BT2 and BT3 sections.
In the BT3, the injection beam asymptotically approaches the stored beam trajectory for the storage ring injection point.

\subsection{Beam injection} \label{subsec:inj_design}
\begin{figure*}[htb]
    \centering
    \includegraphics[width=0.8\linewidth]{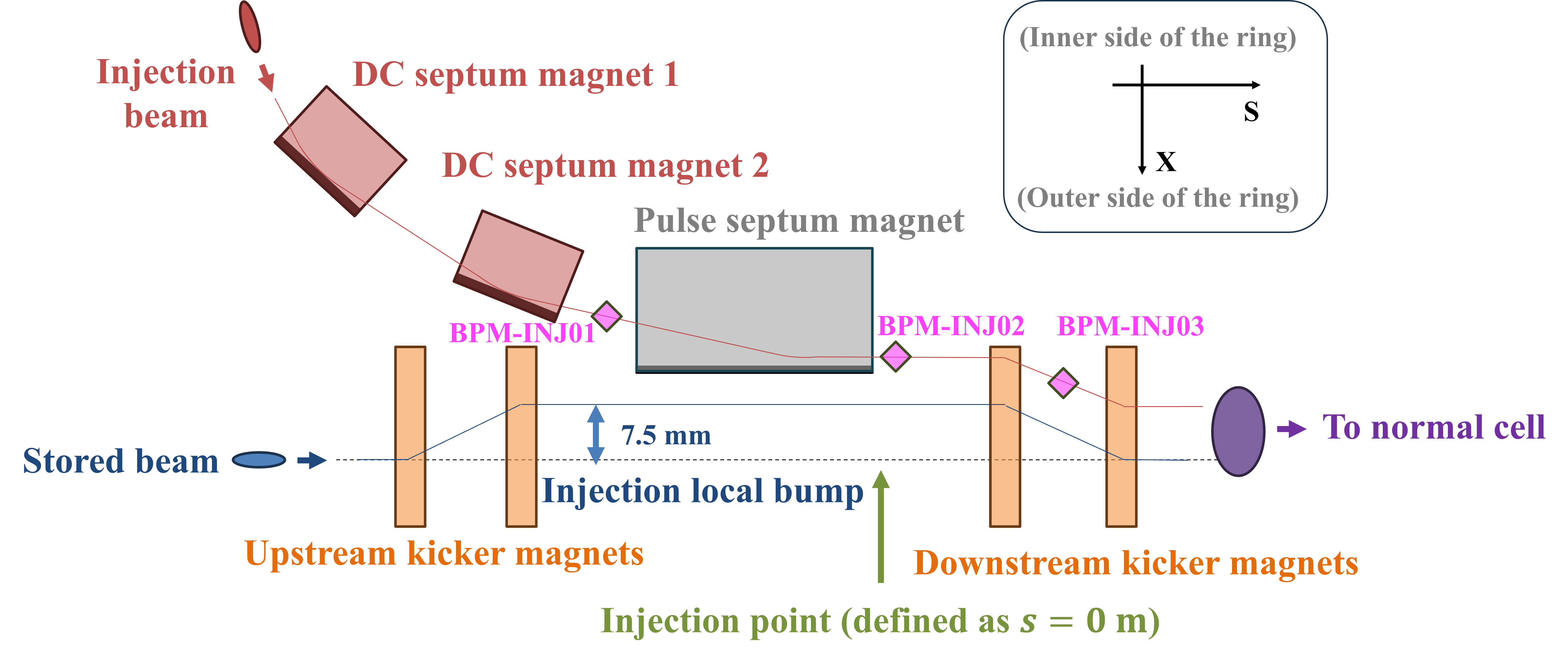}
    \caption{Schematic of the storage ring injection area. 
    A stored beam going from the left side of this figure is kicked by pulse kickers, and a $7.5$-mm-high local bump is produced horizontally. 
    An injection beam from the beam transport line comes from the upper side of this figure and is curved horizontally into the inner side of the ring by two DC septum magnets and a pulse septum magnet.
    The injection beam can be monitored by the three beam-position monitors (BPMs).}
    \label{fig:srinj}
\end{figure*}

An in-vacuum off-axis beam injection system from the ring inside is used for stable and transparent beam injection and to allow top-up operation~\cite{Takano:IPAC2019-WEPMP009}. 
The injection system shown in Fig.~\ref{fig:srinj} consists of two compact DC septum magnets with a recently proposed low-current-density design~\cite{yamaguchi_design_2023}, an in-vacuum pulse septum magnet, and a pair of twin kickers with identical magnetic characteristics thanks to a new iron lamination scheme with additional interlaminar insulation~\cite{FukamiKickerMagnet2022}. 
The system is designed to allow a small oscillation amplitude of the injected electron beam and a stored beam oscillation amplitude less than $10\,\mu{\rm m}$, satisfying transparent beam injection during top-up operation. 
The bumped beam trajectory is $7.5\,{\rm mm}$ from the nominal position horizontally.
The septum wall of thickness $0.5\,{\rm mm}$ is $2.5\,{\rm mm}$ from the bumped stored beam. 
The length and kick angle of each kicker are $300\,{\rm mm}$ and $6\,{\rm mrad}$, respectively.
Those of the pulse septum are $500\,{\rm mm}$ and $70\,{\rm mrad}$, and those of the DC septum are $400\,{\rm mm}$ and $48\,{\rm mrad}$.
The parameters are summarized in Tab.~\ref{tab:tableSrInjParams}.

\begin{table}[htb]
    \caption{\label{tab:tableSrInjParams}%
    Parameters of the storage ring injection magnets~\cite{nanoterasu_cdr}. }
    \begin{ruledtabular}
        \begin{tabular}{lccccc}
        & DC septum & Pulse septum & Kicker \\ \hline
        Magnetic field (T) & 1.2 & 1.4 & 0.2 \\
        Effective length (mm) & 400 & 500 & 300 \\
        Integrated field (Tm) & 0.48 & 0.70 & 0.06 \\
        Kick angle (mrad) & 48.0 & 70.0 & 6.0 \\
        Pulse width ($\mu$s) & - & 10 & 3 \\
        \end{tabular}
    \end{ruledtabular}
\end{table}

Two DC septum magnets are powered in series by a single power supply. 
Twin kickers are driven by a single solid-state pulsar to generate identical kicker magnetic pulses~\cite{Inagaki:IPAC2018-WEYGBF4}. 
The pulsar consists of a charging circuit up to $55\,{\rm kV}$, a main capacitor of $65\,{\rm nF}$, and a fast IGBT switch, and provides a half-sine pulse-shaped current with a $1.6\,{\rm kA}$ peak and 3\,$\mu$s width. 
The fluctuation of the charging voltage is measured as $0.01\%$, indicating good reproducibility. 
The magnetic properties of the kickers, such as inductance, should be identical~\cite{FukamiKickerMagnet2022}. 
The difference between the measured temporal profiles of the magnetic fields of the twin kickers is $\pm 0.1$\%, satisfying the specification, corresponding to $\pm6\,\mu{\rm rad}$ kick uncertainties. 
Ceramic vacuum chambers with uniform titanium coating of 3$\pm 0.1$\,$\mu$m in thickness are used. 
The magnetic field is attenuated by $3\%$ due to the eddy current in the coating, and the attenuation is compensated for by increasing the power supply output. 

The in-vacuum pulse septum magnet has a thin septum wall 0.5\,mm thick to allow the stored and injected beams to get closer for small injected-beam oscillation amplitudes. 
The magnet gap is $2\,{\rm mm}$, which is narrow enough to provide a $1.4\,{\rm T}$ pulsed magnetic field uniformly along the horizontal axis over $4\,{\rm mm}$. 
The stray field integral outside of the septum is reduced to $1\times10^{-5}\,{\rm Tm}$ or less by using a permalloy magnetic shield. 
The vacuum pressure is measured as $\sim 5\times 10^{-8}\,{\rm Pa}$. 
We measured the pulsed magnetic field as a function of horizontal position with a thin search coil. 
The field is flat within $\pm 0.2\%$ from $2$ to $5\,{\rm mm}$ from the septum wall, where the injected beam passes.
The commissioning results of the ring beam injection system are described in Sec.~\ref{subsec:inj_commissioning}.

\subsection{Storage ring} \label{subsec:sr_design}
\begin{figure*}[htb]
    \centering
    \includegraphics[width=\linewidth]{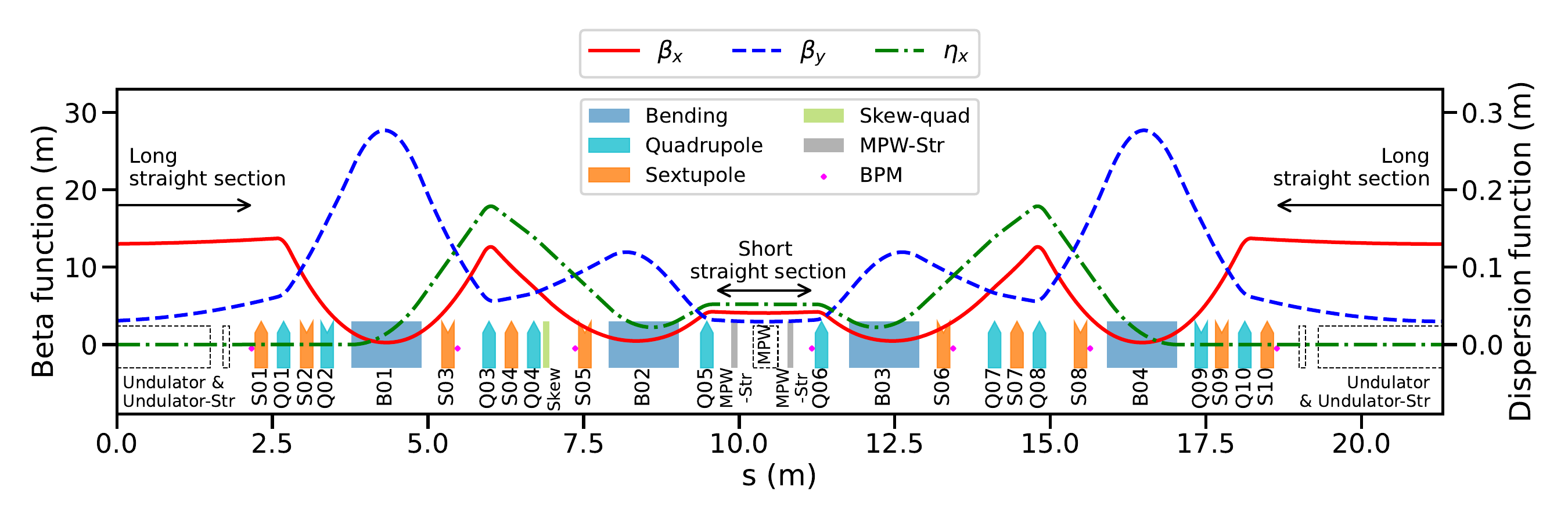}
    \caption{Lattice functions and magnet layout for a cell. 
    The design is symmetrical around the center, which is the MPW installation position. 
    The horizontal axis represents the path length starting from the end of the pulse septum magnet at the storage ring injection area.
    }    
    \label{fig:opticsmodel}
\end{figure*}

\begin{table}[htb]
    \caption{\label{tab:tableSrParams}%
    Main parameters of the NanoTerasu storage ring four-bend achromat lattice~\cite{nanoterasu_cdr}. }
    \begin{ruledtabular}
        \begin{tabular}{lc}
        Beam energy & $2.998\,{\rm GeV}$\\
        Lattice & Four-bend achromat \\
        Circumference length & $348.843\,{\rm m}$ \\
        Number of cells & 16 \\
        Number of bending magnets & 4$\times$16 \\
        Number of quadrupole magnets & 10$\times$16 \\
        Number of sextupole magnets & 10$\times$16\\
        Long straight section for undulator & $5.44\,{\rm m} \times 14$\\
        Short straight section for MPW & $1.64\,{\rm m} \times 14$ \\
        Betatron tune $(\nu_x, \nu_y)$ & (28.17, 9.23) \\
        Natural horizontal emittance $(\varepsilon_x)$ & $1.14\,{\rm nm}\,{\rm rad}$ \\
        Momentum compaction factor ($\alpha$) & $4.3\times10^{-4}$\\    
        Momentum spread ($\sigma_E/E$) & $0.0843\%$ \\
        Bunch length (@ $0\,{\rm mA}$) & $2.92\,{\rm mm}$ \\ 
        Optics at undulator center ($\beta_x$, $\beta_y$, $\eta_x$) & $\qty(13.0\,{\rm m}, 3.0\,{\rm m}, 0.0\,{\rm m})$  \\
        Optics at MPW center $\qty(\beta_x, \beta_y, \eta_x)$ & $\qty(4.1\,{\rm m}, 3.0\,{\rm m}, 0.05\,{\rm m})$ \\ 
        Beam current & $400\,{\rm mA}$\\
        Energy loss in bends & $0.621\,{\rm MeV/turn}$ \\
        Accelerating radio frequency & $508.759\,{\rm MHz}$\\
        Harmonic number & $592$ \\        
        \end{tabular}
    \end{ruledtabular}
\end{table}

The NanoTerasu lattice functions and magnet layouts for a unit cell are shown in Fig.~\ref{fig:opticsmodel}, and the accelerator machine parameters are summarized in Tab.~\ref{tab:tableSrParams}. 
One cell consists of 4 B-Q combined bending magnets, 10 quadrupole magnets (8 focusing and 2 defocusing), 10 sextupole magnets (4 focusing and 6 defocusing), 1 quadrupole skew magnet, and 2 independent steering magnets (MPW-Str).
All magnets are aligned within $50\,\mu{\rm m}$ uncertainties to ensure an optics correction.
Details of the magnet alignment procedures and evaluation using a laser tracker (AT403, Leica) and vibrating wire are discussed in Appendix~\ref{appendix:alignment}.
The main magnet parameters are summarized in Tab.~\ref{tab:MagParams}.
\begin{table*}[htb]
    \caption{\label{tab:MagParams}%
    Main parameters of the NanoTerasu storage ring magnets~\cite{nanoterasu_cdr}.}
    \begin{ruledtabular}
        \begin{tabular}{lccc}
        \textrm{Magnet species} & \textrm{Name} & \textrm{Effective length} & \textrm{Strength}\\ \hline
        B-Q combined dipole & B01, B02, B03, and B04 series & $1.13\,{\rm m}$ & $0.8688\,{\rm T}$ and $-7.06\,{\rm T/m}$\\ 
        \hline
        {}         & Q01/Q10 series & $0.20\,{\rm m}$ & $+32.6151\,{\rm T/m}$\\
        {}         & Q02/Q09 series & $0.20\,{\rm m}$ & $-2.6278\,{\rm T/m}$\\
        Quadrupole & Q03/Q08 series & $0.20\,{\rm m}$ & $+49.1523\,{\rm T/m}$\\
        {}         & Q04/Q07 series & $0.20\,{\rm m}$ & $+5.4300\,{\rm T/m}$\\        
        {}         & Q05/Q06 series & $0.20\,{\rm m}$ & $+44.6423\,{\rm T/m}$\\ 
        \hline
        {}         & S01/S10 series & $0.20\,{\rm m}$ & $+949.50\,{\rm T/m^2}$\\
        {}         & S02/S09 series & $0.20\,{\rm m}$ & $-1124.87\,{\rm T/m^2}$\\
        Sextupole  & S03/S08 series & $0.20\,{\rm m}$ & $-472.34\,{\rm T/m^2}$\\
        {}         & S04/S07 series & $0.20\,{\rm m}$ & $+1061.73\,{\rm T/m^2}$\\        
        {}         & S05/S06 series & $0.20\,{\rm m}$ & $-1540.68\,{\rm T/m^2}$\\  
        \end{tabular}
    \end{ruledtabular}
\end{table*}
The insertion device is placed between the cells, corresponding to the sides of Fig.~\ref{fig:opticsmodel}, and the MPW is at the center of the cell.
The stored beam is monitored by the button-type beam position monitors (BPMs)~\cite{masaki_design_2016}, of which there are seven per cell.

All bending magnets (B01, B02, B03, and B04) are connected to one power supply in series, with no auxiliary power supplies or trim coils~\cite{nanoterasu_cdr}.
These bending magnets also have a defocusing quadrupole magnetic field (B-Q combined) in addition to the dipole field.
The damping partition numbers in the horizontal, vertical, and longitudinal directions are 1.389, 1.0, and 1.611, respectively~\cite{nanoterasu_cdr}.

The quadrupole magnets are categorized into five families (Q01/Q10, Q02/Q09, Q03/Q08, Q04/Q07, and Q05/Q06), and each family is connected to one power supply in series. 
For tune correction, we use one focusing quadrupole family (Q01/Q10) and one defocusing quadrupole family (Q02/Q09) in the non-dispersive sections.
Five of the 10 quadrupole magnets (Q01, Q03, Q06, Q08, and Q10) have independent auxiliary power supplies for optics corrections, mainly horizontal dispersion and beta functions.
The vertical dispersion and horizontal-vertical coupling can be optimized by using the skew magnets.

The sextupole magnets are also categorized into five families (S01/S10, S02/S09, S03/S08, S04/S07, and S05/S06). 
Each family of sextupole magnets is connected to one power supply in series.
For use as steering magnets, six of 10 sextupole magnets (S01, S03, S05, S06, S08, and S10) have independent auxiliary power supplies in each pole pairs. 
Horizontal and/or vertical magnetic fields for steering are realized by the balances of the independent three-pair (six-pole) currents.
Therefore, eight horizontal/vertical steering magnets per cell, including two MPW-steering magnets, can contribute to the closed orbit correction.
The maximum kick angles for the sextupole steering magnet and MPW steering magnet are $0.4$ and $0.2\,{\rm mrad}$, respectively.
In the chromaticity correction, we use focusing (S03/S08) and defocusing (S04/S07) sextupole families in the dispersive section.
The optics corrections are discussed in Sec.~\ref{subsec:codcorrection} and Sec.~\ref{subsec:loco}.

The maximum radiation loss per turn is estimated to be $1.25\,{\rm MeV}$ where $0.62\,{\rm MeV}$ is lost in the bending magnets, and the rest is lost in the insertion devices when an average power of 9\,kW is assumed for 28 insertion devices with a nominal stored beam current of $400\,{\rm mA}$. 
The maximum radiation power is evaluated to be $500\,{\rm kW}$ for a $400\,{\rm mA}$ beam. 
To make the Touschek lifetime longer than 5\,h, an RF accelerating voltage of $3.3\,{\rm MV}$ is required. 
Four sets of $508.76\,{\rm MHz}$ acceleration cavities are installed on one of the 5.4-m-long straight sections. 
A new type of compact TM020-mode normal conducting cavity has been developed to accommodate the space limitation~\cite{Ego:FLS2023-tu3d4,EGO2024169418}. 
The TM020 cavity has magnetic field nodes at the TM020 acceleration mode inside the cavity. 
We prepared slots along the nodes and installed ferrite to damp higher-order modes (HOMs) entering the slots because the HOMs excited in the cavity cause coupled-bunch instability. 
HOMs are dissipated on the ferrite dampers, whereas the TM020 acceleration mode is not damped. 
The shunt impedance of the new cavity is estimated to be $6.8\,{\rm M}\Omega$ and an RF power of $100\,{\rm kW}$ per cavity is required for 0.8\,MV acceleration, which corresponds to an acceleration of $3.3\,{\rm MV}$ and an RF power of $400\,{\rm kW}$ per four cavities.
Thus, a total klystron RF power of $0.9\,{\rm MW}$ is required to provide the maximum radiation power from a $400\,{\rm mA}$ stored beam.

\section{Beam Commissioning} \label{sec:commissioning}
\begin{figure*}[htb]
    \centering
    \includegraphics[width=\linewidth]{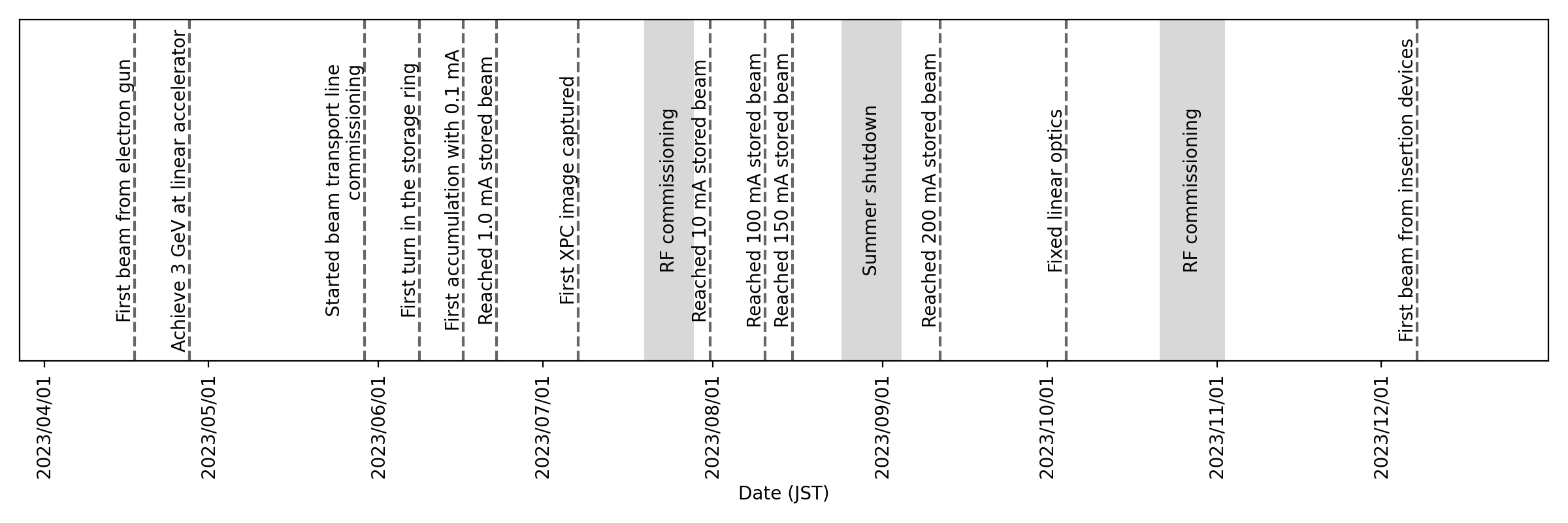}
    \caption{Timeline from April 2023 to December 2023.
    Some important events are shown by vertical dashed lines.
    The gray shaded periods correspond to machine shutdown times.}
    \label{fig:events}   
\end{figure*}
The NanoTerasu buildings were constructed from March 2019 to March 2022 in Sendai, Japan.
Installation of all magnets, vacuum chambers, accelerating RF cavities, BPMs, and power supplies started in December 2021 and finished in January 2023 for the linear accelerator and in May 2023 for the storage ring.
Conditioning of the accelerating cavities of the linear accelerator started on 13 February 2023 while the installation work proceeded in the storage ring tunnel.
After two months, on 17 April 2023, we started extracting electron beams from the gun in the linear accelerator.
Section~\ref{subsec:linac_commissioning} describes the linear accelerator commissioning works up to obtaining a 3\,GeV energy beam as the injector.
The first electron accumulation was achieved on 8 June 2023, as described in Sec.~\ref{subsec:inj_commissioning}.
By commissioning the accelerating RF cavity (Sec.~\ref{subsec:SRRF_commissioning}), vacuum (Sec.~\ref{subsec:srvac_commissioning}), and monitor (Sec.~\ref{subsec:srbpm_commissioning}), we performed the optics corrections (Sec.~\ref{subsec:codcorrection} and Sec.~\ref{subsec:loco}).
Model-consistent ring optics and a $200\,{\rm mA}$ stored beam current were obtained on 4 October 2023 by our half-year machine commissioning since the first electron beam generation by the gun.
The first X-ray beam from the insertion device was obtained on 7 December 2023, and the beamline commissioning was also started in parallel.
Finally, user operation began on 9 April 2024 with a stable 160\,mA stored current, as described in Sec.~\ref{sec:usertime}.
A chronological timeline with the commissioning highlights is shown in Fig.~\ref{fig:events}.
In this section, we describe each accelerator machine commissioning result.

\subsection{Linear accelerator and beam transport line} \label{subsec:linac_commissioning}
The RF conditioning at the linear accelerator started in February 2023, and the first electron beam was generated by the electron gun in April 2023. 
After 10 days, we achieved a beam energy of 3\,GeV with $0.3\,{\rm nC}$ and transportation to a beam dump. 
In May 2023, the beam route was changed in the beam transport line going to the storage ring. 
In June 2023, the beam reached the end of the beam transport section, where a beam profile was measured on a screen monitor between the two DC septum magnets in the storage ring injection area. 

The electron beam generated in the gridded thermionic cathode is accelerated by an electric field between the anode and cathode for an energy of 50\,keV. 
The 50\,keV beam is immediately accelerated to 500\,keV in the 238\,MHz cavity at a crest acceleration phase. 
The accelerating phase and voltage of the 238\,MHz cavity were optimized via a time-of-flight (TOF) measurement using two current transformers (CT) near the SHB cavity, the signals of which were measured by an oscilloscope. 
According to simulation results using {\textsc{PARMELA}}~\cite{Young:PARMELA2003}, the experimental dependence of the TOF signal on the SHB phase and voltage was adjusted. 
After the SHB, the electron beam is accelerated to 40\,MeV in the 2\,m S-band accelerating structure. 
The electron beam is transported through a chicane to the main C-band section. 
The beam-induced field is measured by the RF pickup of each C-band accelerating unit to adjust the accelerating phase and beam position. Finally, the accelerated 3\,GeV beam at the exit of the C-band section is transported to the storage ring via the beam transport line.

\begin{figure}[thb]
    \centering
    \includegraphics[width=\linewidth]{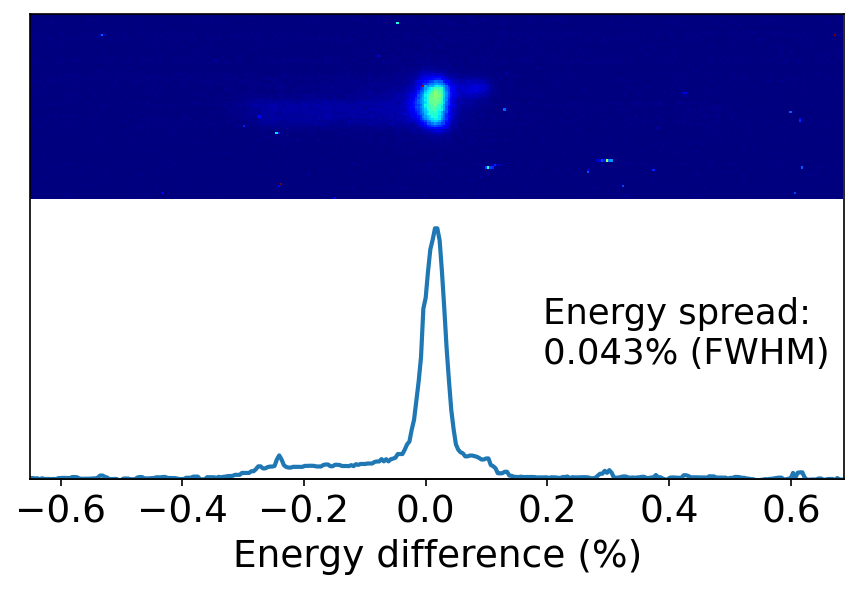}
    \caption{Energy profile of the 3\,GeV electron beam. A screen monitor and bending magnet are used for the measurement at the exit of the linear accelerator.}
    \label{fig:linacEnergyProfile}
\end{figure}

The emittance of the accelerated 3\,GeV electron beam is obtained by the quadrupole scan method~\cite{minty_emittance_2003}, and is approximately 2\,nm\,rad in both the horizontal and vertical directions. 
The stabilities of the bunch charge and energy for the 3\,GeV electron beam are measured as 0.52\% and 0.057\%, respectively, which are stabilized by an RF feedback system. 
Figure~\ref{fig:linacEnergyProfile} shows the energy profile of the 3\,GeV beam measured using a screen monitor at the first dispersive position after the C-band section, which shows an energy spread of 0.043\% (FWHM). 
Calibration was performed by using the dependence of the beam position on the bending magnet currents located upstream of the screen. 
Tab.~\ref{tab:linacParams} summarizes the beam performance during beam commissioning. 
All beam parameters satisfy the designed values (required performance)~\cite{nanoterasu_cdr} for storage ring injection.

\begin{table}[htb]
    \caption{\label{tab:linacParams}%
    Designed values and measurement results for the beam parameters of the 3\,GeV linear accelerator.}
    \begin{ruledtabular}
        \begin{tabular}{lrr}
            \textrm{Items} & \textrm{Design} & \textrm{Measurement} \\ \hline
            Beam energy & 3\,GeV & 3\,GeV\\
            Bunch charge & $>$0.3\,nC & 0.39\,nC \\
            Bunch charge stability & - & 0.52\% \\
            Horizontal emittance & $2\,{\rm nm}\,{\rm rad}$ & $1.04 \pm 0.03 \,{\rm nm}\,{\rm rad}$\\                   
            Vertical emittance & $2\,{\rm nm}\,{\rm rad}$ & $1.84 \pm 0.19 \,{\rm nm}\,{\rm rad}$\\
            Energy spread & 0.16\% (FWHM) & 0.043\% (FWHM) \\
            Energy stability & $<$0.2\% & 0.057\% \\
        \end{tabular}
    \end{ruledtabular}
\end{table}

\subsection{Storage ring injection and first accumulation } \label{subsec:inj_commissioning}

\begin{figure}[htb]
    \centering
    \includegraphics[width=0.9\linewidth]{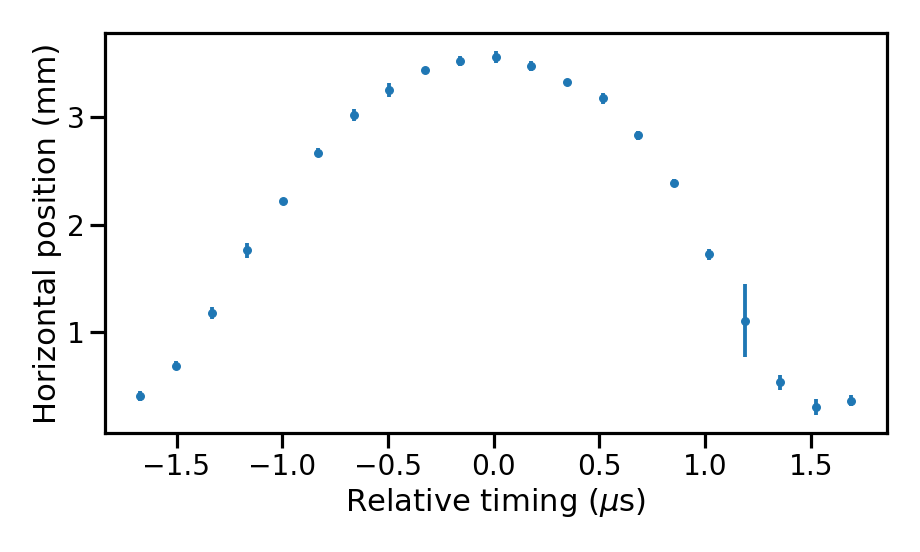} 
    \caption{Injection beam positions as a function of the kicker pulse output timing. The horizontal axis represents the relative timing from the center of the figure set to $0\,\mu{\rm s}$. The error bar also includes the stability of the injection beam position from the linear accelerator.}
    \label{fig:kicker_timing_scan}
\end{figure}

Commissioning of the storage ring beam injection system was performed by using three BPMs (Fig.~\ref{fig:srinj}).
The injection beam trajectory from the beam transport line was adjusted in the designed position by changing the DC septum magnet powers and monitoring with BPM-INJ01.
Because the pulse septum magnet provides a half-sine curve shape output~\cite{Inagaki:IPAC2018-WEYGBF4}, 
the kicked beam position also follows the half-sine curve at BPM-INJ02 if the output timing of the power supply is changed relative to the beam timing.
We measured the half-sine curve beam responses and searched for the flat top timing of the shape.
This method can also be adapted with BPM-INJ03 to the kicker magnets, which use the same type of solid-state power supplies.
Figure~\ref{fig:kicker_timing_scan} shows the horizontal beam position monitored at BPM-INJ03 that was varied by changing the output timing of the downstream kicker power supply.
The kicker power supply outputs a half-sine curve with a width of about $3\,\mu{\rm s}$, and the expected local-bump height at the BPM-INJ03 position is about $3.5\,{\rm mm}$.
The data were consistent with the predictions, and then we optimized the pulse output timings.

The parallelism of the injection beam with respect to the designed stored beam trajectory was tuned using BPM-INJ02 and BPM-INJ03.
Turning off the downstream kickers and turning on only the pulse septum should give the same horizontal beam position between BPM-INJ02 and BPM-INJ03.
We adjusted the pulse septum power as described above.

After optimizing the injection timings and powers, we led the beam into the storage ring.
Owing to the excellent magnet alignment described in Appendix~\ref{appendix:alignment}, the electron beam traveled over $300$ turns in the storage ring with no steering magnet kicks on 8 June 2023.

\begin{figure}[htb]
    \centering
    \includegraphics[width=\linewidth]{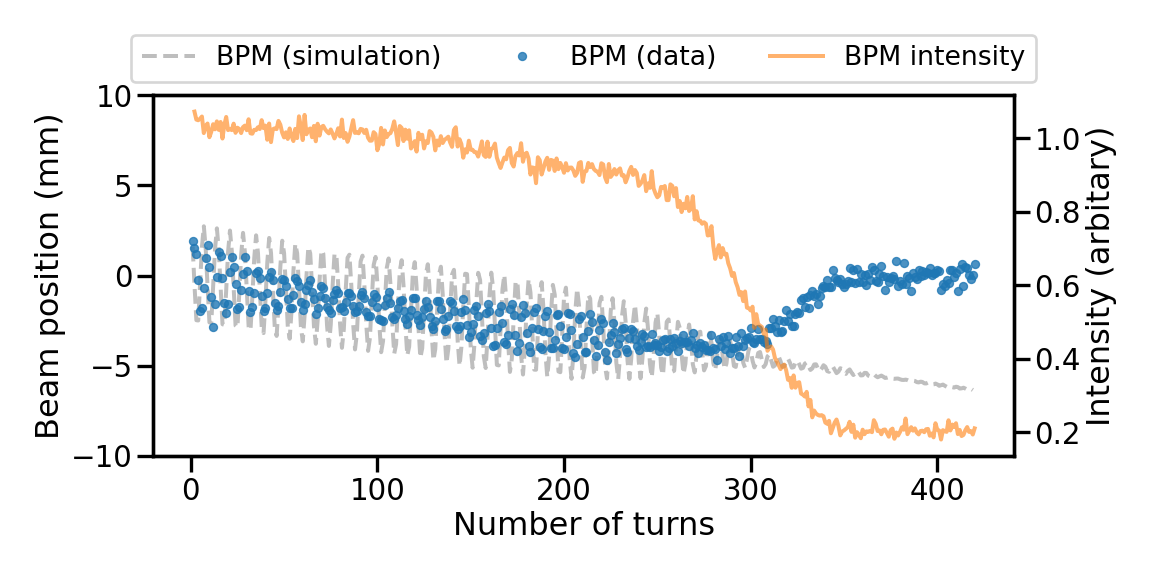}
    \caption{Turn-by-turn horizontal beam position (blue) and BPM intensity (orange) at a dispersive section BPM without RF power. 
    The gray dashed line corresponds to simulated beam positions with ideal injection beam optics.
    In the simulation, $1000$ electrons are tracked with the model storage ring parameters.
    }
    \label{fig:turn300}
\end{figure}

Figure~\ref{fig:turn300} shows a turn-by-turn horizontal beam position at one of the dispersive sections ($\eta_x = 0.09\,{\rm m}$) and BPM intensity with no RF power.
After around $300$ turns, the electrons spread concentrically, and the observed beam position appeared near the center.
At around $330$ turns, the BPM signal intensity reached the background level because the beam was lost.
The measured beam positions and betatron oscillation almost agreed with the simulation, except for the amplitude damping, although the accelerator machines were under commissioning.
It is suspected that these deviations were caused by the incident beam energy and optics.
The beam transport and linear accelerator sections can still be adjusted to address these discrepancies.

We achieved the first capture and storage of an electron beam on 16 June 2023 after turning on the RF power and optimizing its phase.  
The first accumulated beam orbit is discussed in Sec.~\ref{subsec:codcorrection}.
This is the first ring beam acceleration by a TM020 mode cavity to our knowledge, and details of the conditioning of TM020 mode cavities are described in Sec.~\ref{subsec:SRRF_commissioning}.

\begin{figure}[htb] 
    \centering 
    \includegraphics[width=\linewidth]{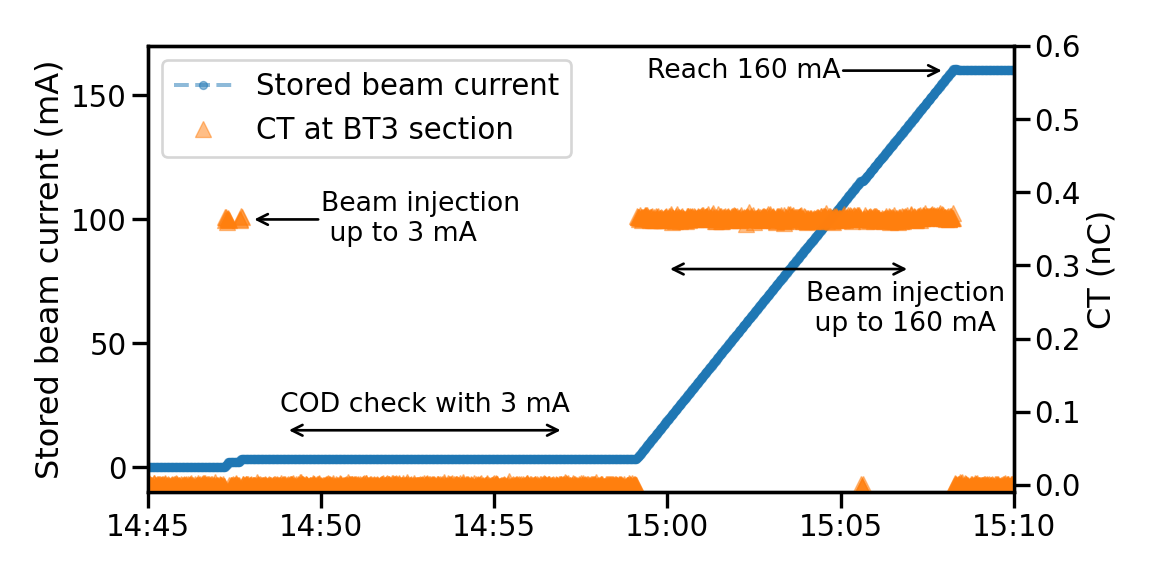} 
    \caption{Stored beam current (blue dots) and injection charge per shot in $1\,{\rm Hz}$ monitored at the BT3 section (orange triangles). } 
    \label{fig:injeff}
\end{figure}

\begin{figure}[htb]
    \centering
    \includegraphics[width=0.95\linewidth]{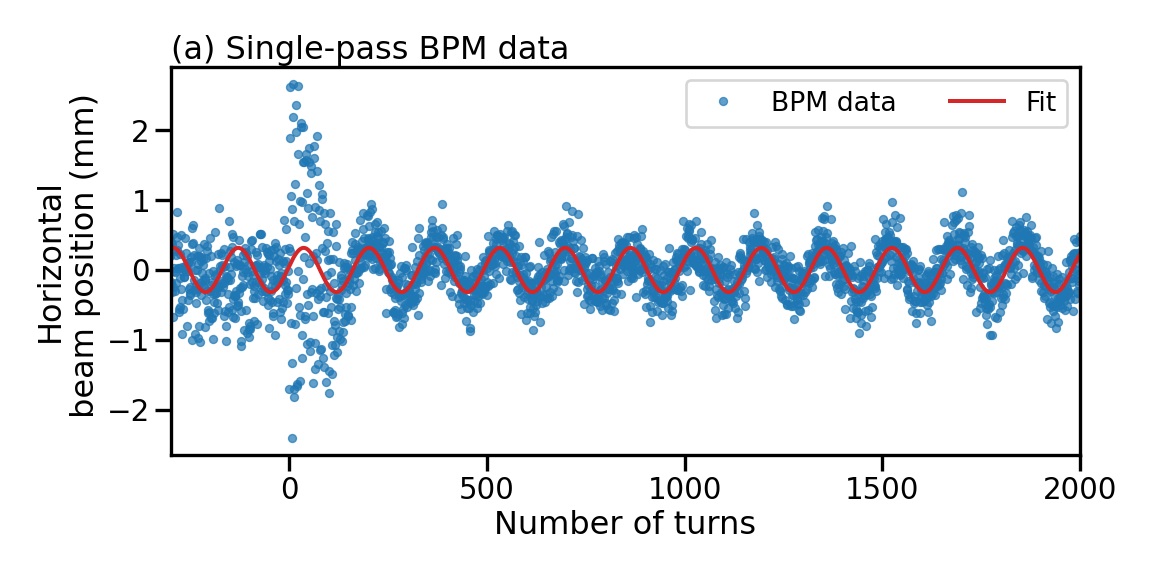}
    \includegraphics[width=0.9\linewidth]{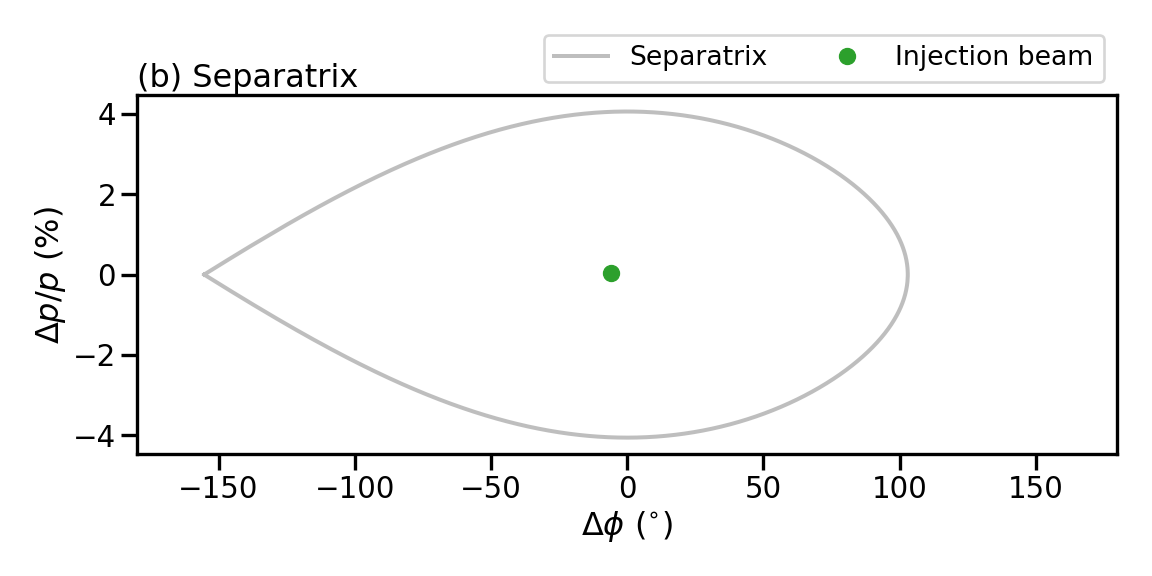}
    \caption{(a) Injection beam position data observed by a single-pass BPM at the dispersive section in the storage ring. 
    The red line represents a fitting result with a cosine function to estimate the beam phase.
    (b) Injection beam synchrotron phase ($\Delta \phi$) and energy differences ($\Delta p/p$). 
    The separatrix is drawn assuming an accelerating voltage of $2.9\,{\rm MV}$ (Sec.~\ref{subsec:SRRF_commissioning}). 
    The injection beam has little energy discrepancy as $\Delta p/p = 0.046\%$ and synchrotron phase as $\Delta \phi = -5.9\,{\rm degree}$.}
    \label{fig:separatrix}
\end{figure}

Owing to the well-established beam transportation in the beam transport line and linear accelerator commissioning, the injection beam charge into the storage ring is $\sim0.3\,{\rm mA}$ per shot, corresponding to an injection efficiency of $\sim88\%$.
One of the usual beam storing processes up to 160\,mA is shown in Fig.~\ref{fig:injeff}.
Figure~\ref{fig:separatrix} shows the injection beam profiles monitored by a BPM at one of the dispersive sections to estimate the synchrotron phase.
The beam position oscillates $\sim2\,{\rm mm}$ at the injection timing but converges to less than $0.5\,{\rm mm}$ after $\sim30$ turns.
In contrast, at the BPM near the undulator positions where there is the dispersion-free section, the converged horizontal oscillation is less than $\pm 0.1\,{\rm mm}$ after $\sim30$ turns.
The beam is injected at the center of the separatrix.

\subsection{Storage ring RF commissioning} \label{subsec:SRRF_commissioning}

\begin{figure}[htb]
    \centering
    \includegraphics[width=\linewidth]{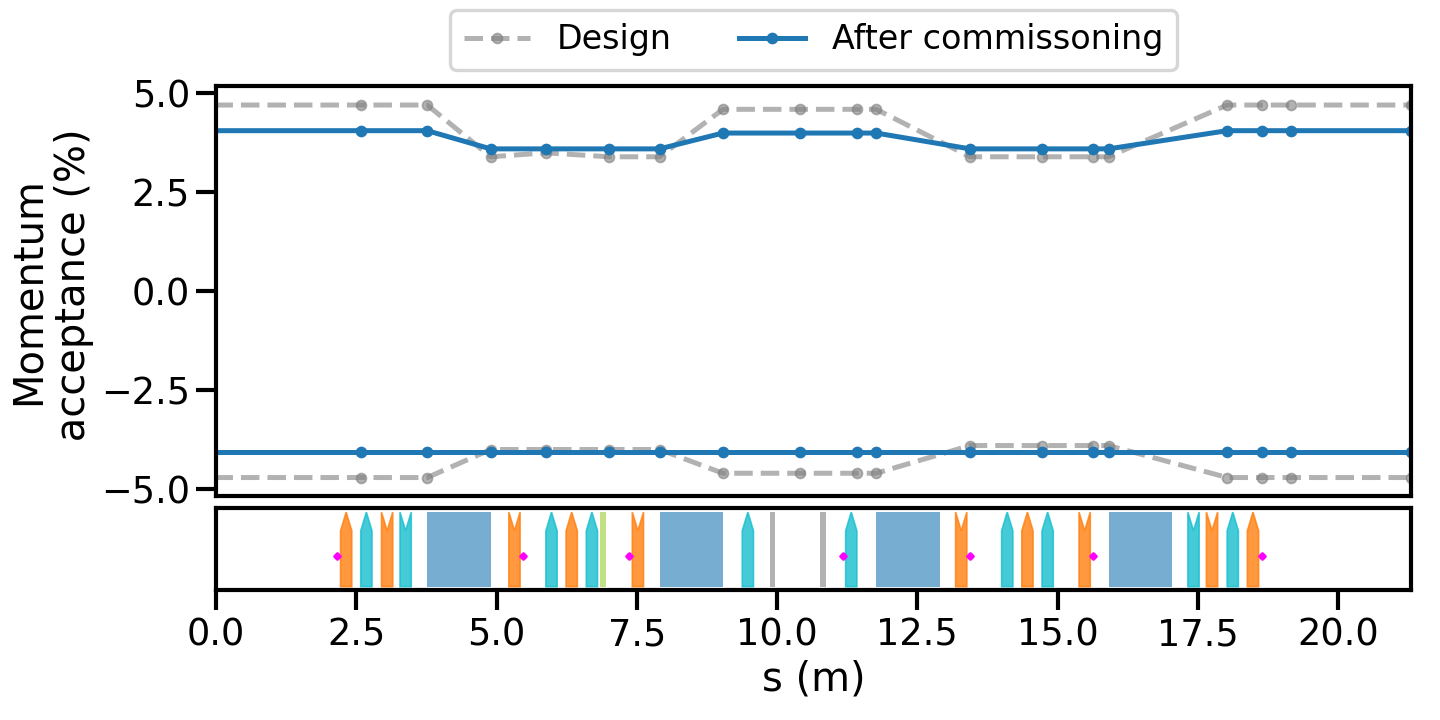}
    \caption{Momentum acceptance estimation in a cell. The gray (blue) line shows the momentum acceptance with a 3.6\,MV (2.9\,MV) accelerating voltage configuration. The bottom panel displays magnet and BPM positions as shown in Fig.~\ref{fig:opticsmodel}.}
    \label{fig:momentum_acceptance}
\end{figure}

As the RF conditioning progressed, we gradually increased the input RF powers of four TM020-mode cavities installed in a long straight section. 
By the time we achieved the first electron beam accumulation on 16 June 2023, the total power of the four cavities had reached $200\,{\rm kW}$, and by mid-July, it had reached a maximum of $400\,{\rm kW}$ without an electron beam.
We set the accelerating voltage and RF power to be 2.9\,MV and 320\,kW, respectively, for the initial phase of user operation with $10$ insertion device beamlines. 
The bunch length spreads to $3.3\,{\rm mm}$ from a designed value of $2.9\,{\rm mm}$ for the accelerating voltage. The RF momentum acceptance is estimated to be $3.7\%$ when $10$ insertion devices are operated with an average power of $9\,{\rm kW}$ (Fig.~\ref{fig:momentum_acceptance}). 
This acceptance is similar to $3.5\%$ evaluated for the $3.2\,{\rm MV}$ accelerating voltage with full operation of $28$ insertion devices discussed in Sec.~\ref{subsec:sr_design}, where a Touschek lifetime longer than 5\,h is anticipated.

\subsection{Storage ring vacuum} \label{subsec:srvac_commissioning}

\begin{figure}[htb]
    \centering
    \includegraphics[width=\linewidth]{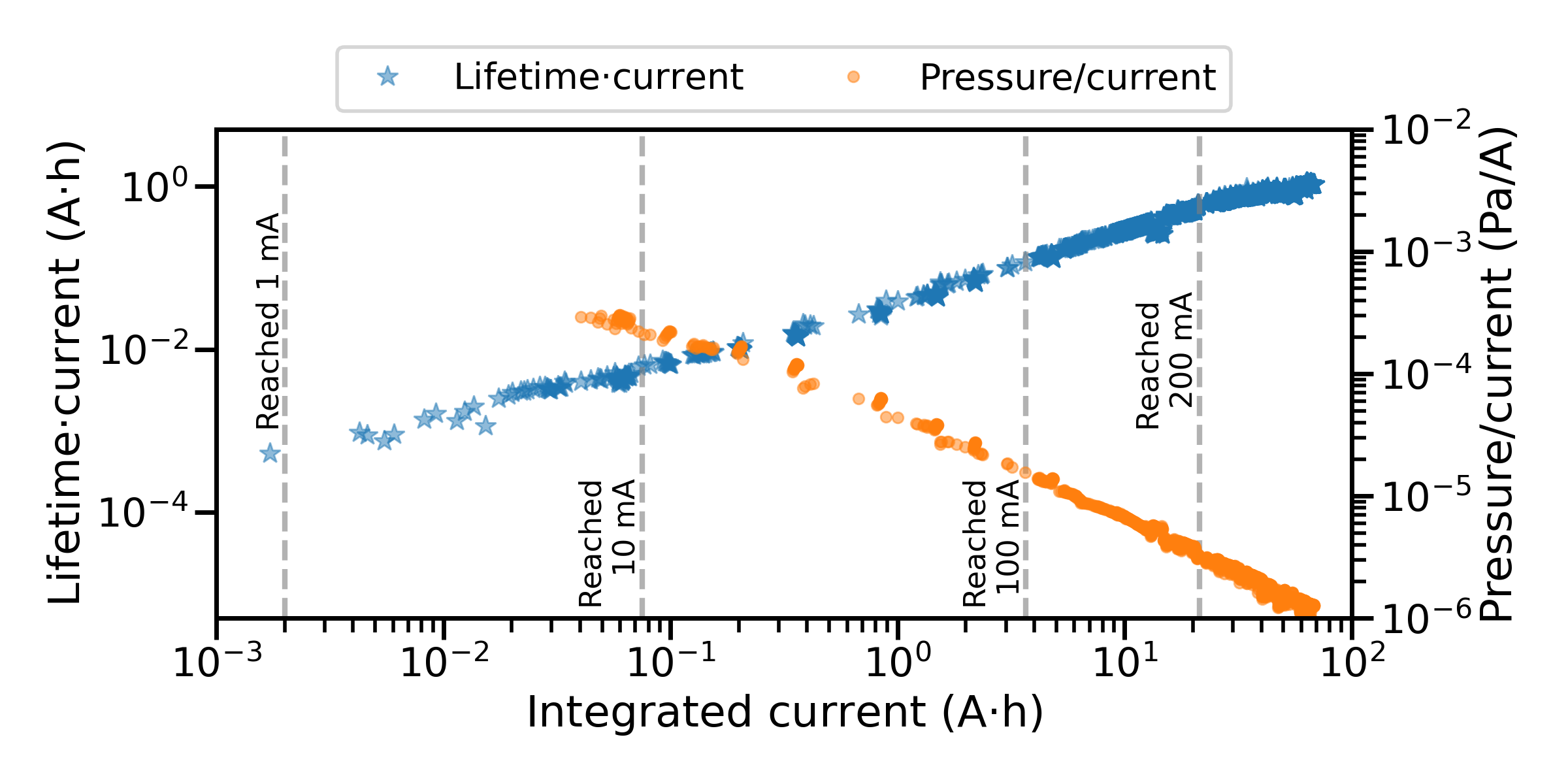}
    \caption{Electron lifetime and ring vacuum pressure for a half year. 
    Lifetime multiplied by the stored beam current grows with respect to an integrated current.
    Pressure normalized by the stored beam current decreases with respect to the integrated current.
    The pressure value is taken from a cold-cathode gauge near the photon beam absorber. 
    The gray vertical dashed lines correspond to some milestones of the maximum stored beam current.}
    \label{fig:dose}   
\end{figure}

Optics corrections and beam monitor commissions began in the daytime with a stored beam current of $10\,{\rm mA}$ or less.
For vacuum conditioning, we started night operation in August 2023 with a stored beam current of over $10\,{\rm mA}$. 
We gradually increased the stored current for the vacuum conditioning, as the maximum stored current was updated in daytime commissioning (Fig.~\ref{fig:events}).
Finally, we performed $200\,{\rm mA}$ stable operation over the whole night.
Figure~\ref{fig:dose} shows the progress of the vacuum conditioning in the storage ring.
We obtained a beam lifetime of over 5\,h with a stored beam configuration of $200\,{\rm mA}$ until December 2023.
Moreover, by continuing the vacuum conditioning while commissioning the accelerator and beamlines, the electron beam lifetime reached $\sim10$\,h, which is a sufficiently long time for the first year of user operation.

\subsection{Storage ring BPM} \label{subsec:srbpm_commissioning}

\begin{figure}[htb]
    \centering
    \includegraphics[width=\linewidth]{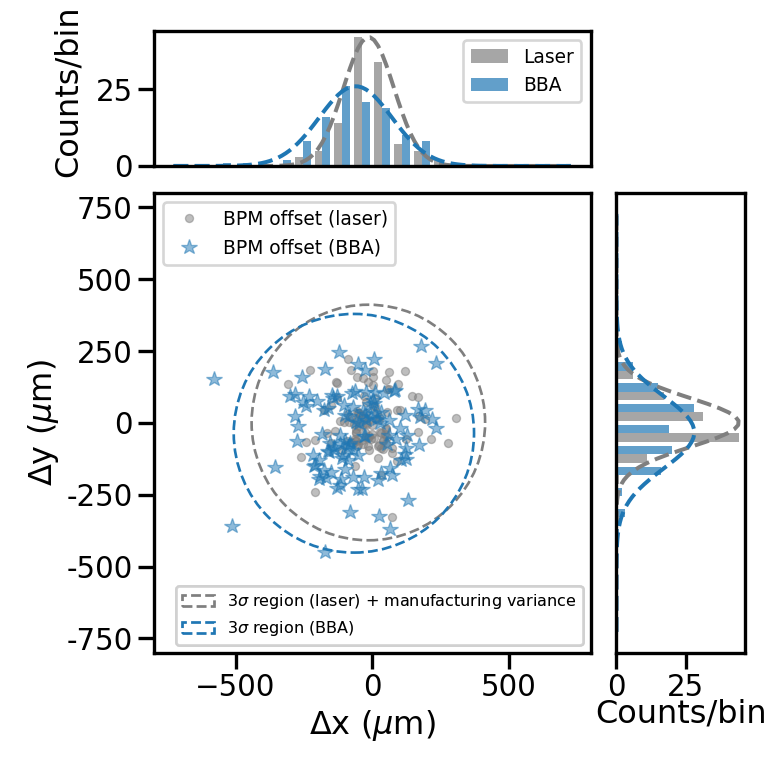}
    \caption{BPM offsets from the BBA and laser tracker measurements.
    The top and right panels represent the one-dimensional projected histograms horizontally and vertically.
    Dashed curves in the top and right panels are Gaussian fitting results.
    The blue dashed ellipse in the left bottom panel is a 3-$\sigma$ region from the Gaussian fitting results.
    The gray dashed ellipse is a 3-$\sigma$ region from the Gaussian fitting results, considering the manufacturing uncertainty among the BPM electrodes and laser tracker basepoint.}
    \label{fig:bba}   
\end{figure}
Beam positions are monitored by seven BPMs per cell (Fig.~\ref{fig:opticsmodel}).
These BPMs are mounted on the vacuum chambers; therefore, there is little range in their independent alignments.
The vacuum chambers were aligned by the laser tracker measurement, with priority on the alignment of the photon beam absorber positions.
The laser tracker accuracy is better than 50\,$\mu$m in our configuration, but the manufacturing uncertainty between the mechanical and electrical centers of BPMs is about $100\,\mu{\rm m}$.
To ensure the precise beam position, the beam-based alignment (BBA) method~\cite{PhysRevAccelBeams.23.012802} is essential.
Figure~\ref{fig:bba} shows the BPM offsets obtained by BBA.
The most distant horizontal and vertical offsets are $-578$ and $-450\,\mu{\rm m}$, respectively.
The mean value and standard deviation of laser tracker measurements for horizontal (vertical) are $-13.89$ ($1.38$) and $101.46$ ($93.10\,\mu{\rm m}$), respectively.
Those of the BBA are -66.82 (-36.07) and 146.61 (138.42\,$\mu$m), respectively.

Hereafter, beam positions are corrected by using the above results.

\subsection{Closed orbit and tune correction} \label{subsec:codcorrection}

To tune the ring optics to the designed model, we performed the following steps: closed orbit distortion (COD) correction, circumference length correction, betatron tune correction, rough horizontal dispersion correction, chromaticity correction, and dispersion and beta function correction using a linear optics correction.
This section describes the optics corrections except for the beta function correction. 
In the next section, we discuss the ring optics and error sources with dispersion and beta function correction.

\begin{figure}[htb]
    \centering
    \includegraphics[width=\linewidth]{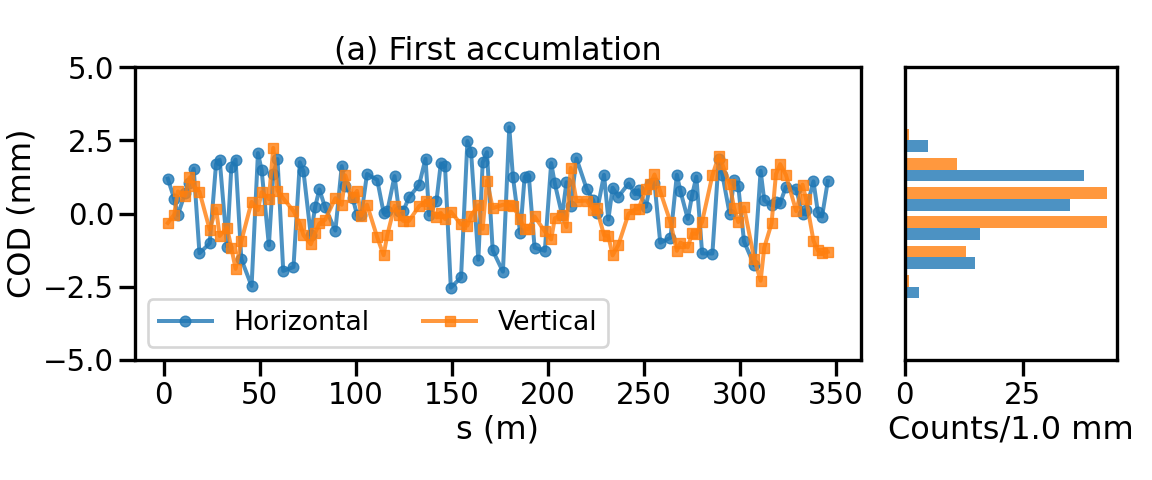} 
    \includegraphics[width=\linewidth]{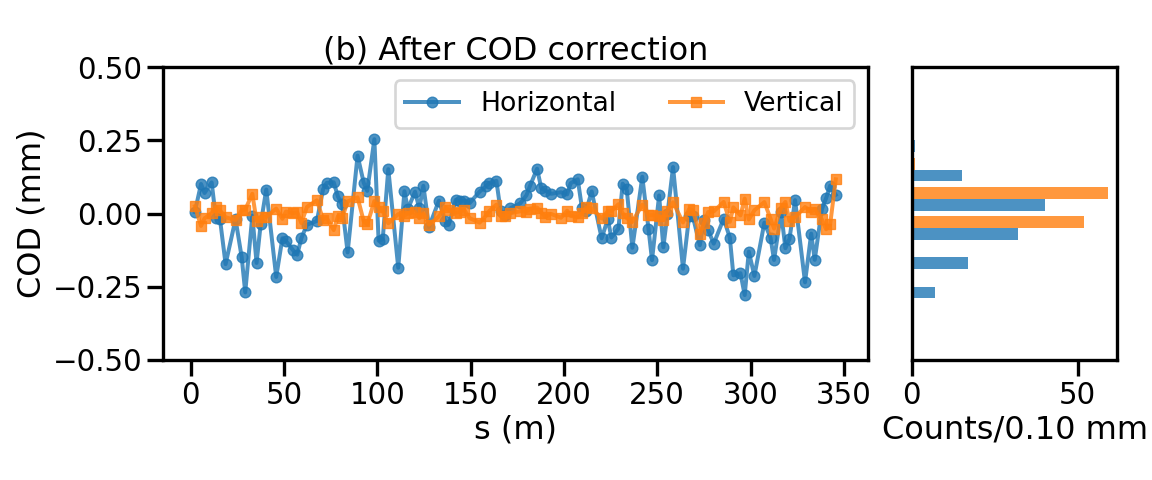}
    \caption{Closed orbit at the first electron beam accumulation in the storage ring (top panel), and after COD and other optics corrections (bottom panel), monitored by the 112 BPMs in the storage ring. 
    Top panel: The horizontal (vertical) mean values and standard deviations are $0.44$ ($-0.01$), and $1.17$ ($0.84\,{\rm mm}$), respectively.
    Bottom panel: The horizontal (vertical) mean values and standard deviations are $-0.01$ ($0.00$), and $0.11$ ($0.03\,{\rm mm}$), respectively.}
    \label{fig:cod}
\end{figure}

Figure~\ref{fig:cod}-(a) shows the (COD) at the first electron beam accumulation on 16 June 2023.
The first COD was a vertical spread of $\sim\pm 2\,{\rm mm}$ compared with the vacuum chamber height of $\pm 8\,{\rm mm}$ for the normal cell and $\pm 4.5\,{\rm mm}$ for the insertion device sections.
The chamber width is $\pm 15\,{\rm mm}$, which was wide enough for the first horizontal spread of $\pm 3\,{\rm mm}$.
Details of the vacuum chamber are described in the conceptual design report~\cite{nanoterasu_cdr}.

The mean of the horizontal COD was about $0.44\,{\rm mm}$, which corresponds to a longer circumference length of $\Delta C = 1.5\,{\rm mm}$ than the designed orbit.
To modify the difference in circumference, we changed the RF according to
\begin{eqnarray}
    \frac{\Delta f_{\rm rev}}{f_{\rm rev}} &=& - \qty( \alpha - \frac{1}{\gamma^2}) \frac{\Delta p}{p},\\
    \frac{\Delta C}{C} &=& \alpha \qty(\frac{\Delta p}{p}),
\end{eqnarray}
where $f_{\rm rev}$ is the revolution frequency, $\alpha$ is the momentum compaction factor, $\gamma$ is the Lorentz factor, $p$ is the momentum of the synchronous particle, and $C$ is the circumference length.

\begin{figure}[htb]
    \centering
    \includegraphics[width=\linewidth]{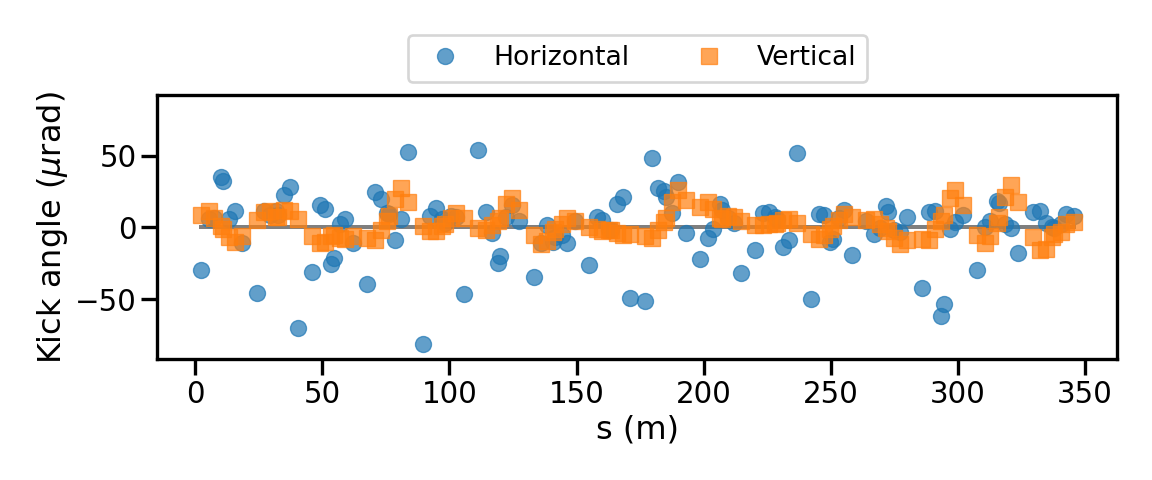} 
    \caption{Horizontal and vertical kick angles applied by all steering magnets.}
    \label{fig:str}
\end{figure}

To correct the COD, we used the steering magnets.
Assuming the kick response function ($R_{m,n}$) for BPM $m$ and steering magnet $n$, a kicked COD ($x^{\rm COD}_{m}$) at BPM $m$ follows:
\begin{eqnarray}
    x^{\rm COD}_m &=& \sum_n R_{m,n} \theta_{n},\\
    R_{m,n} &=& \frac{\sqrt{\beta_m \beta_n}}{2\sin{\qty(\pi \nu)}} \cos{\qty( \abs{\psi_m - \psi_n} - \pi \nu)} \label{eq:cod_response},
\end{eqnarray}
where $\qty(\beta_m, \psi_m)$ are the beta function and the betatron phase at the $m$-th BPM position, $\qty(\beta_n, \psi_n)$ are the beta function and the betatron phase at the $n$-th steering magnet position, $\theta_n$ is the kick angle of the $n$-th steering magnet, and $\nu$ is the betatron tune.
We have 112 BPMs as monitors ($\va{x} = \qty(x_1^{\rm COD}, x_2^{\rm COD}, \cdots, x_n^{\rm COD}, \cdots, x_{112}^{\rm COD})^T$) and 128 horizontal/vertical steering magnets ($\va{\theta} = \qty(\theta_1, \theta_2, \cdots, \theta_n, \cdots, \theta_{128})^T$).
By solving the equation 
\begin{eqnarray}
    \va{x} &=& \vb{M} \cdot \va{\theta},\\
    \vb{M} &\equiv& 
    \mqty( 
    R_{1,1}   & R_{1,2}   & \cdots & R_{1,n}   & \cdots & R_{1,128} \\
    R_{2,1}   & R_{2,2}   & \cdots & R_{2,n}   & \cdots & R_{2,128} \\
    \vdots    & ~         & \ddots & \vdots    & ~      & \vdots \\
    R_{m,1}   & R_{m,2}   & \dots  & R_{m,n}   & \cdots & R_{m,128} \\
    \vdots    & ~         & ~      & \vdots    & \ddots & \vdots \\
    R_{112,1} & R_{112,2} & \cdots & R_{112,n} & \cdots & R_{112,128} 
    ), \label{eq:dipoleresponse}
\end{eqnarray}
we obtained the correction dipole kick angles $\va{\theta}$ adjusting into the center orbit from the observed data $\va{x}$.
We used {\textsc{Eigen}} version~3.4~\cite{eigenweb} for solving the matrices by singular value decomposition.
Figure~\ref{fig:str} shows the dipole kick powers horizontally and vertically for the COD corrections.
The biggest kick was $\sim 80\,\mu{\rm rad}$, which is sufficiently small for the power supplies, which have a maximum designed power of $400\,\mu{\rm rad}$~\cite{nanoterasu_cdr}.
Figure~\ref{fig:cod}-(b) shows the COD after the corrections.
The deviations became $1/10$ both horizontally and vertically, and
the mean values were centered.

\begin{figure}[htb]
    \centering
    \includegraphics[width=\linewidth]{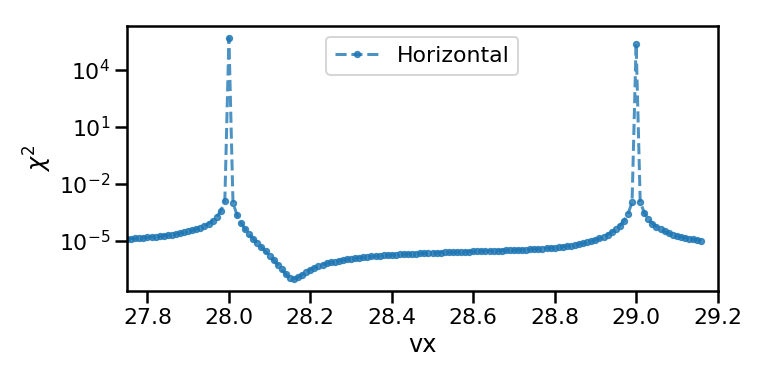}
    \includegraphics[width=\linewidth]{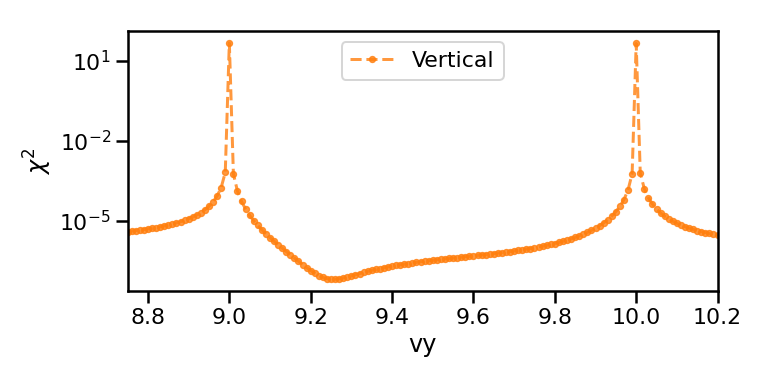}
    \caption{Tune estimation using kicked COD changes for horizontal (upper panel) and vertical (lower panel) tune.}
    \label{fig:estimate_tune}
\end{figure}

The betatron tune is particularly important.
We can monitor the fractional betatron tune using the bunch-by-bunch feedback.
However, in case that the tune is a half-integer different from the designed value, the steering magnet responses work in the opposite direction.
To confirm the tune, including the integer part, we use the beam responses with Eq.~\ref{eq:cod_response}.
A steering magnet applied a small kick to the beam, and we estimated the tune by fitting the COD differences with the model.
Figure~\ref{fig:estimate_tune} shows the fitting results.
The evaluation function is defined as
\begin{equation}
    \chi^2 \equiv \sum_m \qty( \Delta x_m^{\rm COD} - \Delta x_m^{\rm Model}(\nu) )^2,
\end{equation}
where $\Delta x_m^{\rm Model}(\nu)$ is the COD differences as a function of the betatron tune with a steering kick of the $m$-th magnet calculated from the model.
The estimated horizontal tune ($\nu_x$) and vertical tune ($\nu_y$) are $28.16$ and $9.25$, respectively.
These tunes were corrected to $28.17$ and $9.23$ as the designed values by changes in one of the focusing series (Q1/Q10 series) and defocusing series (Q02/Q09 series) of the quadrupole magnets of
\begin{equation}
    \mqty( \Delta \nu_{x} \\ \Delta \nu_{y}) = \frac{e}{4\pi p} \mqty( \beta_{x,F} & \beta_{x, D} \\ -\beta_{y,F} & -\beta_{y,D}) \mqty( k_F L_F \\ k_D L_D ),
\end{equation}
where $\beta_{x(y)}$ is the horizontal (vertical) beta function at the quadrupole magnet position, and subscript $F(D)$ means a focusing (defocusing) series quadrupole magnet.

The dispersion function ($\eta$) was measured by changing the RF to $\pm 300\,{\rm Hz}$;
\begin{equation}
    \eta = \frac{ x^{-300\,{\rm Hz}} - x^{+300\,{\rm Hz}} }{ \qty(\frac{\Delta p}{p})^{-300\,{\rm Hz}} - \qty(\frac{\Delta p}{p})^{+300\,{\rm Hz}}}.
\end{equation}
In our configuration, the momentum difference ($\Delta p/p$) for $\pm 300\,{\rm Hz}$ is $-0.002721$.
The horizontal dispersion function ($\eta_x$) is corrected by tuning the quadrupole magnets, which are aligned at the dispersive section.
We have five auxiliary-connected quadrupole magnets per cell as a free parameter of this optimization, with a total of 80 magnets.
We solved the equation $\va{\eta_x} = \vb{M}_{\eta,x} \va{\theta}_{\eta,x}$ with a response matrix of
\begin{eqnarray}
    \va{\eta}_x &=& \qty(\eta_1, \eta_2, \cdots, \eta_{112})^T,\\
    \va{\theta}_{\eta,x} &=& \qty(k_1 L_1, k_2 L_2, \cdots k_{80} L_{80})^T,\\
    \vb{M}_{\eta,x} &=& \mqty(\ddots & ~ & ~ \\ ~ & R_{m,n} & ~ \\ ~ & ~ & \ddots), \\
    R_{m,n} &\equiv& - \frac{\sqrt{\beta_{x,m} \beta_{x,n}}}{2\sin{\qty(\pi \nu_x)}} \eta_{x,n} \cos{\qty(\pi \nu_x - \abs{\Delta \psi}  )},\\
    \Delta \psi &\equiv& \psi_{x,m}-\psi_{x,n}.
\end{eqnarray}
Those correction quadrupole magnets are also involved in the beta function correction.
We modified only the horizontal dispersion function in the first step.

In contrast, the vertical dispersion is produced by quadrupole-magnet rotation errors and the horizontal dispersion.
We solved the following equation and set the magnets accordingly:
\begin{equation}
    \eta_{y,m} = \sum_n^{N_{\rm skew}} \frac{\sqrt{\beta_{y,m} \beta_{y,n}}}{2\sin{\qty(\pi \nu_y)}} \eta_{x,n} \cos{\qty(\pi \nu_y - \abs{\Delta \psi})} \tilde{k}_n L_n, 
\end{equation}
where $\tilde{k}$ is the skew-quadrupole magnet kick, and ${N}_{\rm skew}$ is the number of skew magnets.

The horizontal and vertical chromaticities $\qty(\xi_x, \xi_y)$, are measured at the timing of the dispersion measurement with RF cavity frequency changes as
\begin{equation}
    \Delta \nu = \xi \frac{\Delta p}{p}.
\end{equation}
The original vertical chromaticity was close to $0$, and the stored beam was unstable at the higher current.
The chromaticity correction was performed with the focusing (S03/S08 series) and defocusing (S04/S07 series) sextupole magnet as
\begin{equation}
    \mqty( \Delta \xi_x \\ \Delta \xi_y) = \frac{e}{2\pi p} \mqty( \eta_{x,F} \beta_{x,F} & \eta_{x,D} \beta_{x,D} \\ -\eta_{x,F} \beta_{y,F} & -\eta_{x,D} \beta_{y,D}) \mqty( \lambda_F L_F \\ \lambda_D L_D  ),
\end{equation}
where $\eta_{x(y)}$ is the horizontal (vertical) dispersion function at the magnet position, subscript $F(D)$ means the focusing (defocusing) sextupole magnet, and $\lambda$ is the sextupole magnet kick. 
For stable operation in the commissioning phase and the first user operation, the modified chromaticities are adjusted to slightly higher values, $\qty(\xi_x, \xi_y) = \qty(1.98, 1.98)$.
Although the present configuration has higher chromaticity than the conceptual design, the beam injection efficiency was not different from that of a configuration fitting the designed value.

\subsection{Linear optics correction} \label{subsec:loco}

In Sec.~\ref{subsec:codcorrection}, we performed the optics corrections in individual steps independently.
These processes are essential to close the storage ring responses in the designed model. 
However, the changes in the quadrupole magnets for tune corrections and dispersion corrections affect the changes in the beta functions.
This section discusses the simultaneous correction for the dispersion and beta function.

Using a fundamental idea of linear optics correction~\cite{SAFRANEK199727}, we measured all response functions (Eq.~\ref{eq:dipoleresponse}) with each steering magnet applying a $30\,\mu{\rm rad}$ kick and estimated the ring optics with errors for the storage ring.
With no error sources in the ring, the responses at the $i$-th BPM of the $j$-th steering kick can be described as,
\begin{eqnarray} 
    \mqty(\Delta x \\ \Delta y)_i &=& \mqty( R^{hh}_{ij} & 0 \\ 0 & R^{vv}_{ij}) \mqty( \Delta \theta_x \\ \Delta \theta_y)_j,\label{eq:locoresponse_noerror} \\
    &=& {\vb{M}} \mqty( \Delta \theta_x \\ \Delta \theta_y)_j, \label{eq:locomatrix_noerror}
\end{eqnarray}
where $R^{hh}$ is the response from the horizontal kick into the horizontal position, and $R^{vv}$ is the response from the vertical kick into the vertical position.
In this case, which corresponds to no errors in our storage ring, the beta function would be the same as the designed value.
Here, we assume some error sources and compute the response functions with the errors.
Comparing the computed and measured responses, we estimate plausible error sources and obtain the optics parameters under those conditions.

Considering the quadrupole magnet kick ($\Delta k$) as an error source, ${\displaystyle \sum_n^{N_{\rm quad}}\pdv{R^{hh(vv)}_{ij}}{k_n}}\Delta k_n$ is added to the diagonal terms.
The rotation error sources of the quadrupole magnets are also considered with additional terms at the anti-diagonal of ${\displaystyle \sum_n^{N} R^{h}_{in}R^{v}_{nj}}\Delta \hat{k}_n$, where the $R^{h(v)}_{in}$ is the skew response function which is horizontally (vertically) kicked by the $n$-th magnet and is monitored by the $i$-th BPM.
The computed matrix (Eq.~\ref{eq:locomatrix_noerror}) can be modified with errors as
\begin{equation} \label{eq:locomatrix_quaderror}
    \vu{M} = \mqty( 
    R^{hh}_{ij} + {\displaystyle \sum_n^{N}\pdv{R^{hh}_{ij}}{k_n}} \Delta k_n & 
    {\displaystyle \sum_n^{N} R^{h}_{in}R^{v}_{nj}}\Delta \hat{k}_n \\
    {\displaystyle \sum_n^{N} R^{v}_{in}R^{h}_{nj}}\Delta \hat{k}_n &
    R^{vv}_{ij} + {\displaystyle \sum_n^{N}\pdv{R^{vv}_{ij}}{k_n}} \Delta k_n).
\end{equation}
The rolling errors of the BPMs and/or the steering magnets with $\Delta \phi$ angles can be written as the rotation matrix, ${\displaystyle \mqty(\cos{\Delta \phi} & -\sin{\Delta \phi}\\ \sin{\Delta \phi} & \cos{\Delta \phi})}$, and the error of their gain powers ($\Delta G$) is expressed as $1+\Delta G$ in the diagonal terms.
With the first-order approximation, their rolling and gain error matrix, $\vb{M}^{rg}$, can be expressed as
\begin{eqnarray} \label{eq:locomatrix_rollgain}
    \vb{M}^{rg} \simeq \mqty( 1+\Delta G_x & -\Delta \phi \\ \Delta \phi & 1+\Delta G_y).
\end{eqnarray}
From Eq.~\ref{eq:locoresponse_noerror},~\ref{eq:locomatrix_quaderror}, and~\ref{eq:locomatrix_rollgain}, the equation for monitoring at the $i$-th BPM and kicking by the $j$-th steering magnet including the errors is
\begin{equation} \label{eq:loco_equation}
    \mqty(\Delta x \\ \Delta y)_i = \vb{M}^{rg}_i \cdot \vu{M} \cdot \vb{M}^{rg}_j \cdot \mqty( \Delta \theta_x \\ \Delta \theta_y)_j.
\end{equation}
To modify the horizontal dispersion, we can consider a similar equation to Eq.~\ref{eq:loco_equation} in which the replacement $\Delta x \rightarrow \Delta \eta_x$ is made.
By solving the equations simultaneously, we obtained the modified values of the error sources.

\begin{figure}[htb]
    \centering
    \includegraphics[width=0.8\linewidth]{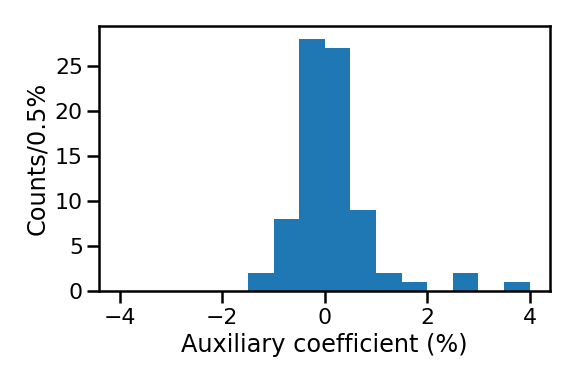}
    \caption{Relative quadrupole magnet fields from the designed values, which the auxiliary power supplies used for the optics correction. Although one magnet was adjusted by a maximum of $+4\%$, the mean of the correction factor is $0.15\%$.} 
    \label{fig:qaux}
\end{figure}

For easy confirmation of the ring responses in the first-year operation, we only modified the quadrupole magnet fields.
The correction coefficients obtained for the auxiliary power supplies are shown in Fig.~\ref{fig:qaux}.

\begin{figure}[htb]
    \centering
    \includegraphics[width=\linewidth]{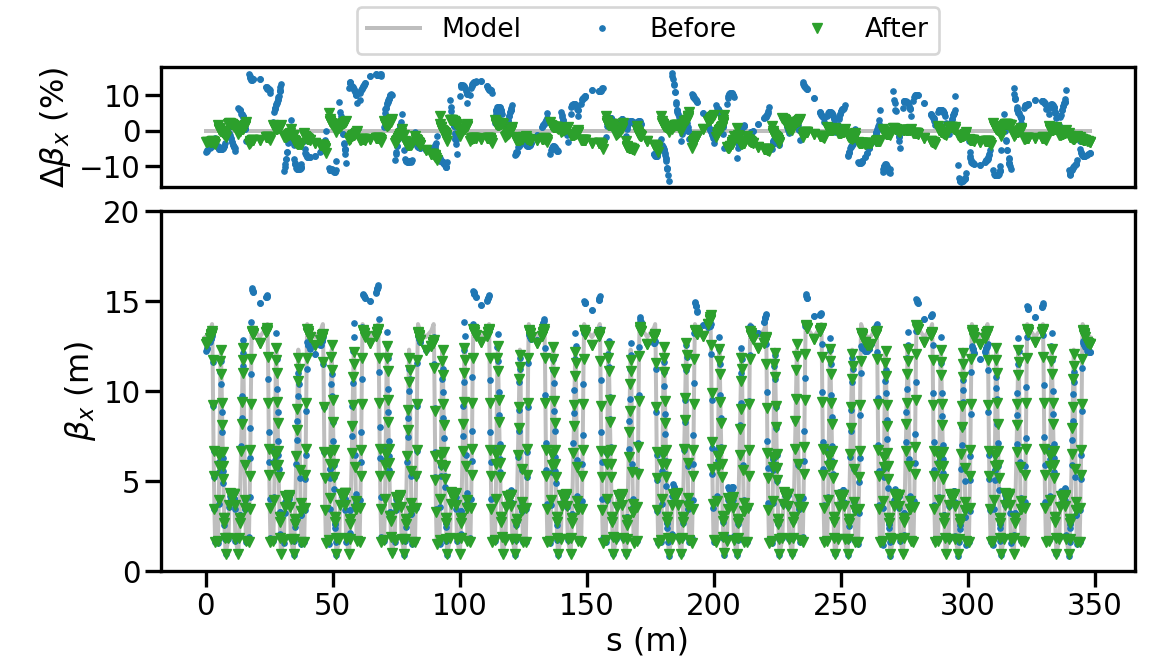}
    \includegraphics[width=\linewidth]{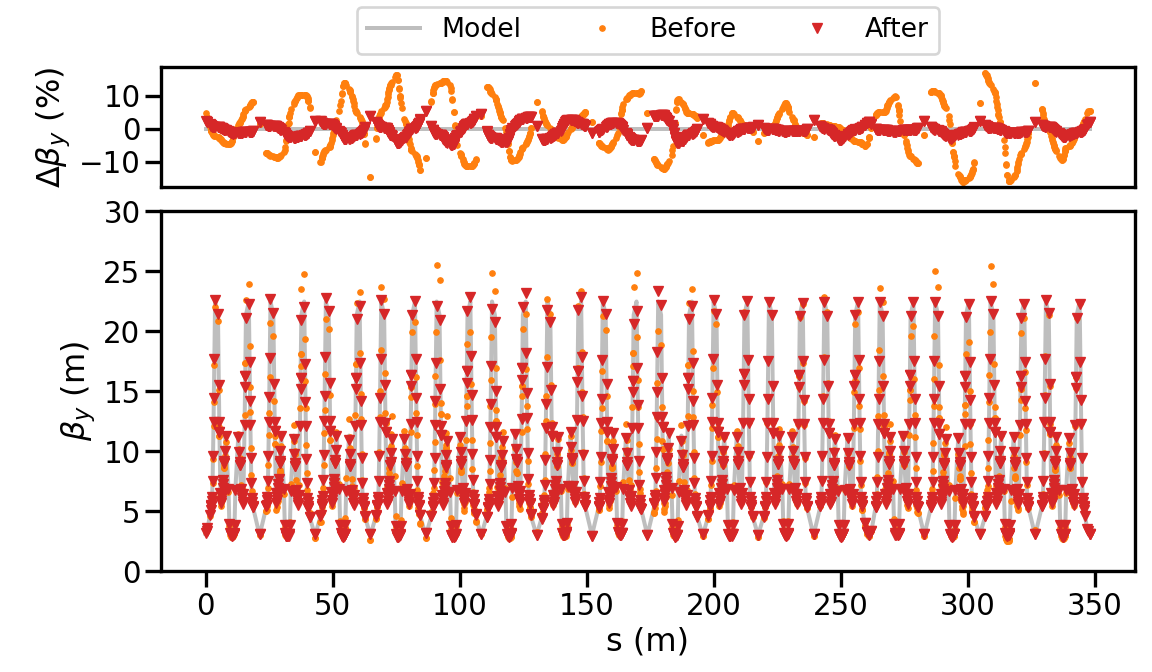}
    \caption{Horizontal (upper panels) and vertical (lower panels) beta functions before and after correction. 
    The top panel of each figure shows the discrepancy from the designed model.}
    \label{fig:dbeta}
\end{figure}

\begin{figure}[htb]
    \centering
    \includegraphics[width=\linewidth]{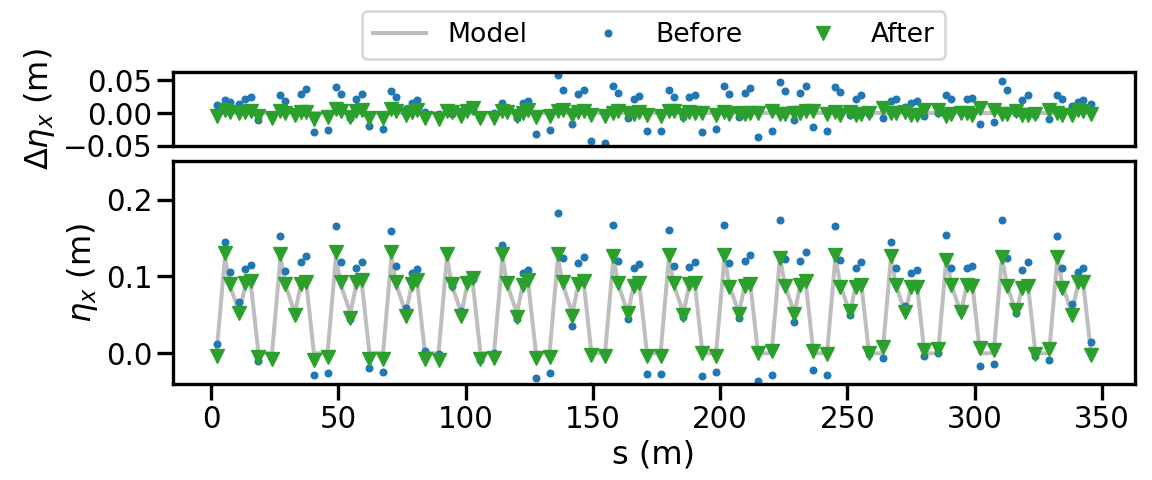}
    \includegraphics[width=\linewidth]{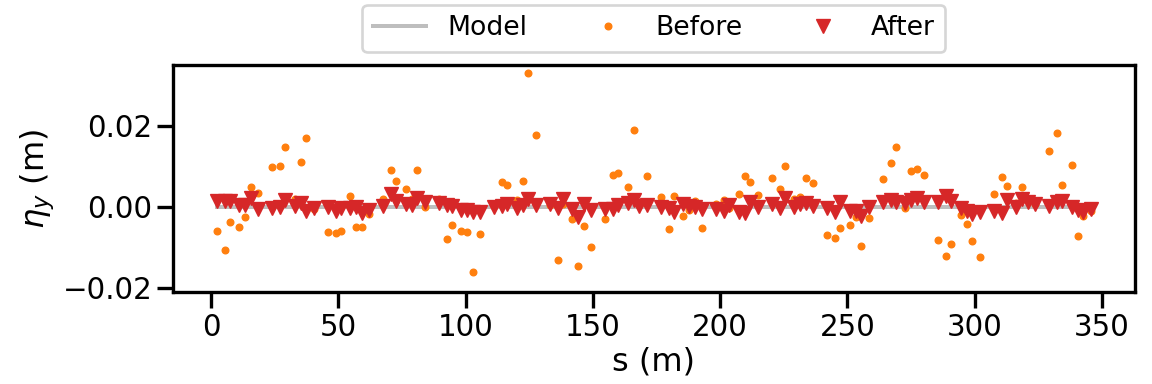}
    \caption{Horizontal (upper panels) and vertical (lower panel) dispersion functions before and after correction. 
    The top panel in the horizontal dispersion shows the difference from the model. 
    Because the vertical dispersion is $0\,{\rm m}$ in the design, the bottom panel corresponds to the difference.}
    \label{fig:dispersion}
\end{figure}

Figure~\ref{fig:dbeta} compares the beta functions before and after the correction and the model.
Although the beta functions were a maximum of $\sim2.0\,{\rm m}$ different from the model before the quadrupole power error corrections, they became consistent within $\sim 0.2\,{\rm m}$ from the design after the modifications.
The horizontal dispersion functions were also modified (Fig.~\ref{fig:dispersion}). 
The discrepancy of the horizontal dispersion was $\pm 0.05\,{\rm m}$ before the correction, but it became consistent within $\pm 0.009\,{\rm m}$.
The vertical dispersion was then independently corrected by using skew magnets, as described in Sec.~\ref{subsec:codcorrection}. 
The vertical dispersion function had a difference of $\pm 0.03\,{\rm m}$ and was modified within $\pm 0.003\,{\rm m}$.

As a result, the discrepancies of the horizontal and vertical dispersion and beta functions reached $1/10$, and our storage ring satisfied the optics parameters sufficiently to start the first-year operation as a synchrotron facility.
We applied the modification only to the quadrupole magnet and not to the BPMs and steering magnets for their rolling and gains, and thus the remaining discrepancies probably originated from these sources. 
Their gain (roll) errors are estimated to be a maximum of approximately $\pm 10\%$ ($\pm4\%$).
We plan to continue optimizing the storage ring step-by-step during machine maintenance time soon.

\subsection{Beam size} 

We have a 3-pole wiggler installed in one of the short straight sections to monitor the beam size.
The wiggler radiation is observed by an X-ray pinhole camera (XPC).
This beam-size monitoring system was developed for the SPring-8-II project~\cite{spring8-II}, and details are given in Ref.~\cite{Takano:IBIC2015-TUCLA02}.
The spatial resolution of the monitoring systems is $\sim 4\,\mu{\rm m}$ for $50\,{\rm keV}$ X-ray conditions~\cite{Ueshima:Pasj2023}.
In our commissioning configuration with different X-ray energy spectra from the study, the net resolution of the monitoring system is $\sim6\,\mu{\rm m}$.

\begin{figure}[htb]
    \includegraphics[width=\linewidth]{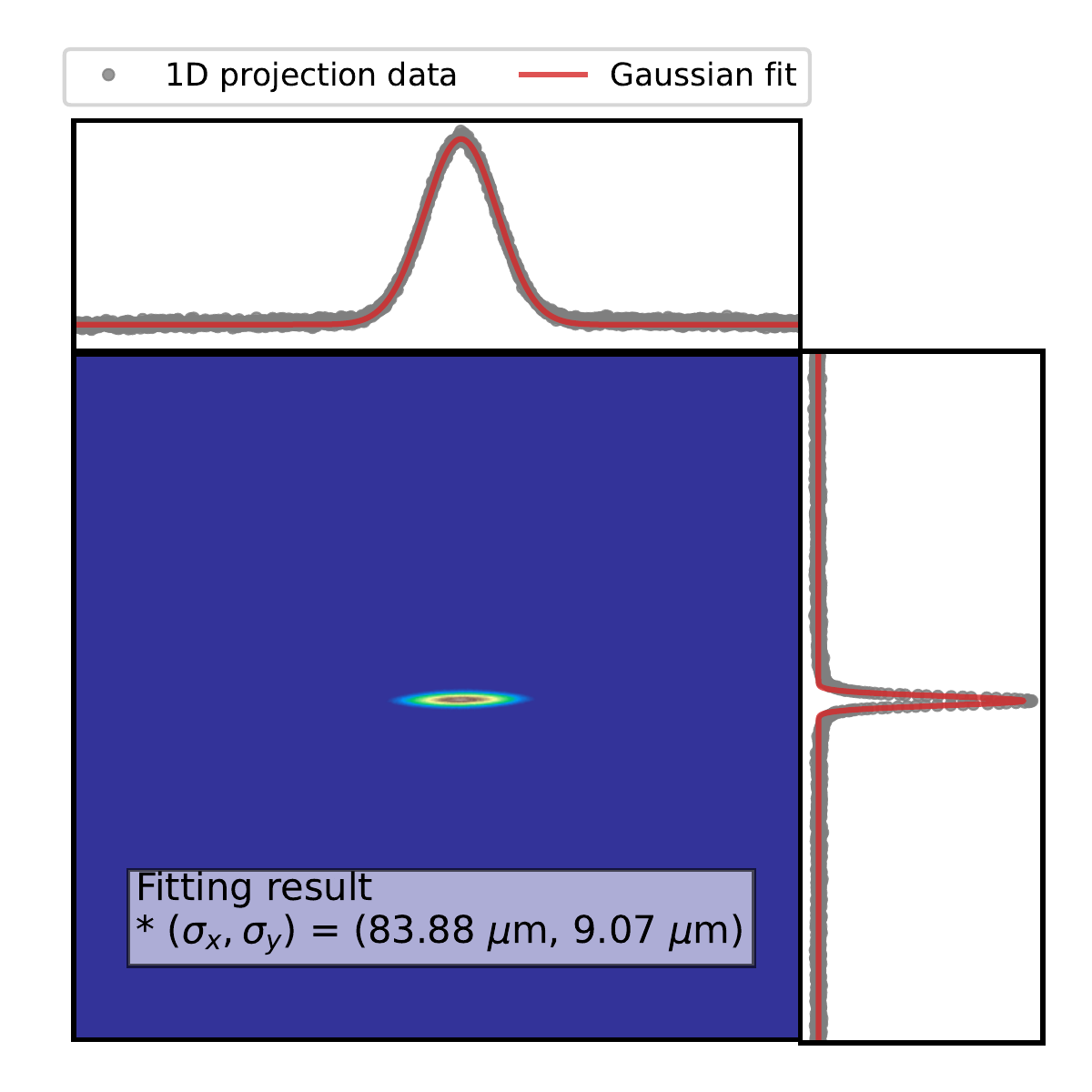}
    \caption{Stored electron beam profile obtained from the XPC with a stored beam current of $10\,{\rm mA}$. 
    Horizontal and vertical beam sizes are estimated from the Gaussian fitting for the one-dimensional projection distributions. 
    These figure axes show the XPC pixels, but the fitting results include the conversion into the actual size of $0.96\,\mu{\rm m}/{\rm pixel}$. 
    The horizontal and vertical axes are arbitrary in this figure and do not directly reflect the beam position in the storage ring.
    }
    \label{fig:xpc_fit}
\end{figure}

Figure~\ref{fig:xpc_fit} is the beam profile monitored by the 3-pole wiggler and XPC.
By projecting into one dimension and fitting with the Gaussian function, we found that the horizontal and the vertical beam sizes are $83.9$ and $9.1\,\mu{\rm m}$, respectively.
The emittance ($\varepsilon$) and energy spread $\qty(\sigma_E/E)$ are calculated by
\begin{equation} \label{eq:emittance_beamsize}    
    \sigma_{\rm beam} = \sqrt{ \qty(\eta \cdot \frac{\sigma_E}{E})^2 + \beta \cdot \varepsilon },
\end{equation}
where $\sigma_{\rm beam}$ is the $1$-$\sigma$ beam size and ($\eta$, $\beta$) are the optics parameters at the 3-pole wiggler position.
We obtained the energy spread as $\qty(\sigma_E/E) = \qty(0.0972\pm0.0161)\%$ assuming a model emittance of $1.14\,{\rm nm}\,{\rm rad}$, that is $1\,\sigma$ consistent within the designed value of $0.084\%$.
The included uncertainties are the discrepancies of the beta function and dispersion between the designed model and the measurement after the correction (Figs.~\ref{fig:dbeta},~\ref{fig:dispersion}).
In contrast, assuming the designed value of the energy spread, the horizontal and vertical emittances are $1.29$ and $0.03\,{\rm nm}\,{\rm rad}$ at the 3-pole wiggler position, respectively.
The coupling between the horizontal and vertical emittance is estimated to be $\sim2.1\%$.
Because the emittance and energy spread are convoluted,  
they cannot be determined as unique values from the XPC results alone.
A future challenge is to use the insertion devices to confirm the compensation between them~\cite{PhysRevAccelBeams.24.081601}.

\section{First User Operation} \label{sec:usertime}

On 9 April 2024 10:00 (JST), the first user time began at NanoTerasu.
The stored current was kept at 159--160\,mA by top-up operation, and the undulators kept their gaps working while the beam was injected.
The beam lifetime is approximately 10\,h, and the beam injection is performed a few times with a 1\,Hz injection every several minutes.
During user operations, we automatically perform the COD correction, circumference length correction, and tune correction (Sec.~\ref{subsec:codcorrection}).
Counter kicks for undulator workings are corrected by feed-forward correction, in which the discrepancies from the normal orbit are less than $\sim \pm 10\,\mu{\rm m}$.

\begin{figure}[htb]
    \includegraphics[width=\linewidth]{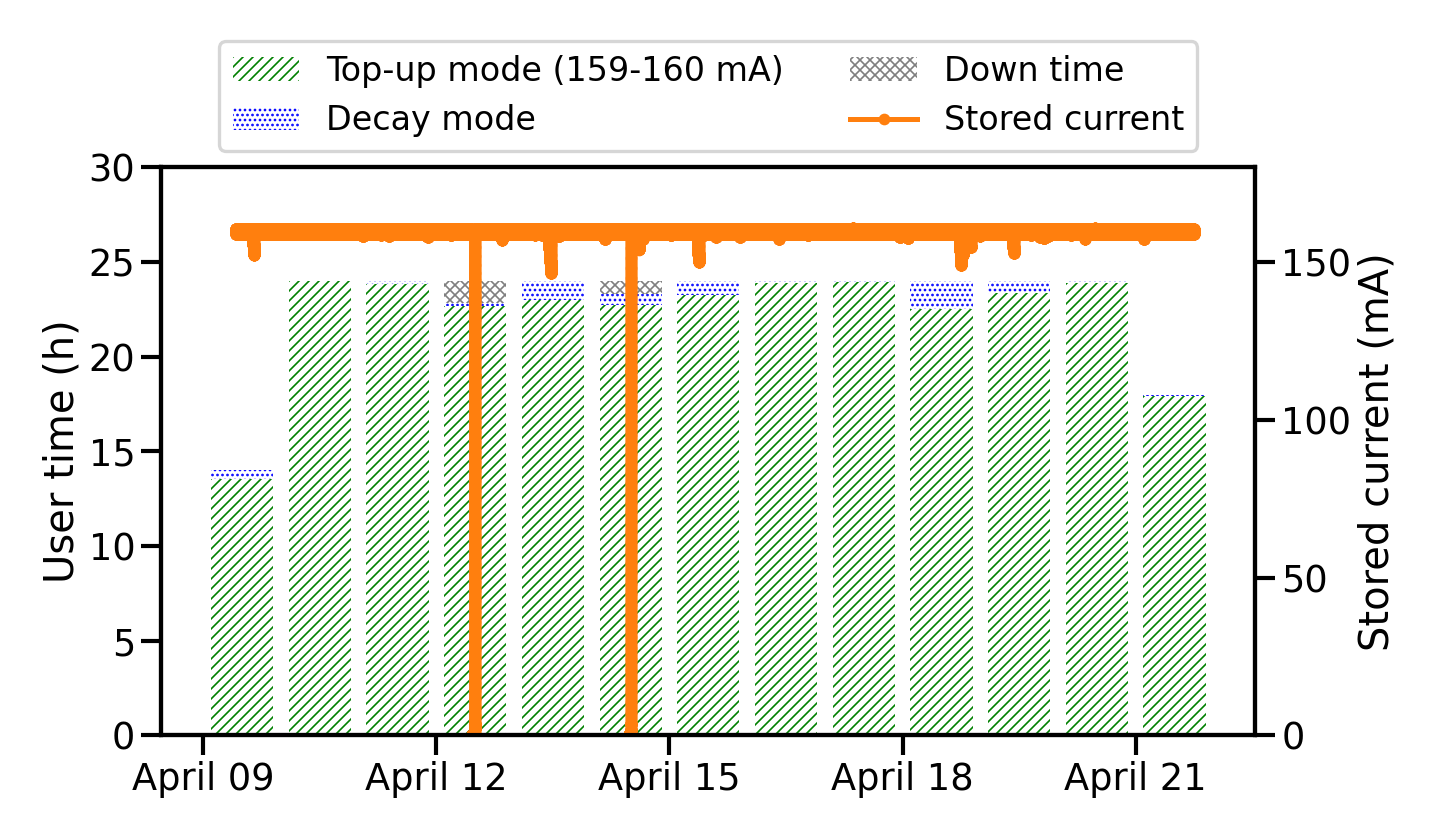}
    \caption{User time and downtime in the first user operation period from 9 April 2024 10:00 to 21 April 2024 18:00 (JST).}
    \label{fig:usertime}
\end{figure}

Figure~\ref{fig:usertime} shows the history of the stored current in the user operation period and user service times.
Recent operation status can be viewed on the NanoTerasu web page~\cite{nanoterasu_web}. 
Until the end of the first user time on 21 April 2024 at 18:00, we experienced beam losses twice due to problems with the storage ring RF cavity reflection.
Because rebooting the RF power to the operational value takes about 1\,h, the total downtime was about 2\,h, corresponding to 0.6\% of the total user time in the first period.
RF power reflections in the linear accelerator also occurred, and the top-up operation was suspended for several minutes several times.
Thus, we achieved 99.4\% user service time in the first period and maintained top-up operation (stored current is more than 159\,mA) in 98.0\% of the time; the assigned user service time was 296\,h, and the total downtime was 1.87\,h.

\section{Summary and Future Prospects} \label{sec:summary}

As shown in the timeline (Fig.~\ref{fig:events}), we achieved a stored beam current of $200\,{\rm mA}$ on 11 September 2023.
In that configuration, we can provide a model-consistent storage ring for user operation.
For user operation from April 2024, we started with a stored beam current of $160\,{\rm mA}$, which is $80\%$ of our stable configuration of $200\,{\rm mA}$.

The storage-ring RF accelerating voltage of $2.9\,{\rm MV}$ is lower than the designed value of $3.6\,{\rm MV}$ but is adequate for the present number of beamlines.
To build new beamlines in the future, we must increase the RF power and perform conditioning on that power.

The coupling constant was not tuned during the present commissioning period but converged with approximately double the designed value.
This will be improved by using skew magnets and the closest difference resonance of $\nu_x - \nu_y = 19$, although we must consider the balance between the Touschek beam lifetime and the vertical emittance, which will be required by beamline users in the future.

The chromaticities are set to be slightly larger than the designed values and will be modified to the designed values while monitoring the beam stability at a higher stored current.

The key point of the short-timeframe commissioning in the storage ring optics was the fine-aligned magnets in the storage ring.
These magnets allowed us to omit the first-turn steering and easily converge the ring optics into the linear beam responses.

The machine commissioning results for the half a year are summarized in Tab.~\ref{tab:summary}.
We achieved model-consistent ring optics, and those parameters will help design a future undulator beamline in NanoTerasu using the \textsc{SPECTRA} simulation tool~\cite{spectra_ttanaka}.

\begin{table*}[hbt]
    \caption{\label{tab:summary}%
    Brief summary of the storage ring commissioning. 
    The beam energy and emittances are assumed to be the designed values.
    $x$ and $y$ represent the horizontal and vertical, respectively.
    }
    \begin{ruledtabular}
        \begin{tabular}{lll}
        \textrm{Parameter} & \textrm{Design} & \textrm{After commissioning} \\ \hline
        Energy & $2.998\,{\rm GeV}$ & $2.998\,{\rm GeV}$\footnotemark[1] \\
        Circumference length $\qty(C)$ & $348.843\,{\rm m}$ & $348.843\,{\rm m}$\\
        Beam current & $400\,{\rm mA}$ & $\geqq 200\,{\rm mA}$\\
        Betatron tune $\qty(\nu_x, \nu_y)$ & $\qty(28.17, 9.23)$ & $\qty(28.17\pm0.01, 9.23\pm0.01)$\footnotemark[2]\\
        Chromaticity $\qty(\xi_x,\xi_y)$ & $\qty(1.38, 1.53)$ & $\qty(1.98, 1.98)$ \\
        Beta function $\qty(\beta_x, \beta_y)$ at the long straight section & $\qty(13.0\,{\rm m}, 3.0\,{\rm m})$ & $\qty(13.0\pm0.3\,{\rm m}, 3.0\pm0.05\,{\rm m})$ \\
        Dispersion $\qty(\eta_x, \eta_y)$ at the long straight section & $\qty(0.0\,{\rm m}, 0.0\,{\rm m})$ & $\qty(0.0\pm0.004\,{\rm m}, 0.0\pm0.001\,{\rm m})$ \\
        Beta function $\qty(\beta_x, \beta_y)$ at the short straight section & $\qty(4.1\,{\rm m}, 3.0\,{\rm m})$ & $\qty(4.1\pm0.1\,{\rm m}, 3.0\pm0.05\,{\rm m})$ \\
        Dispersion $\qty(\eta_x, \eta_y)$ at the short straight section & $\qty(0.05\,{\rm m}, 0.00\,{\rm m})$ & $\qty(0.05\pm0.004\,{\rm m}, 0.00\pm0.001\,{\rm m})$ \\
        COD $\qty(\Delta x, \Delta y)$ &  -- & $\qty(-0.01\pm0.11\,{\rm mm}, 0.00\pm0.03\,{\rm mm})$\footnotemark[2]\\
        Coupling & $1\%$ & $2.1\%$ \\
        Emittance $\qty(\epsilon_x, \epsilon_y)$ & $\qty(1.14\,{\rm nm}\,{\rm rad}, 0.01\,{\rm nm}\,{\rm rad})$ & $\qty(1.14\,{\rm nm}\,{\rm rad}\footnotemark[1]\footnotemark[3], 0.02\,{\rm nm}\,{\rm rad}\footnotemark[4])$\\        
        Energy spread $\qty(\sigma_E/E)$ & $0.0843\%$ & $0.0972\pm0.0161\%$ \\
        Bunch length $\qty(\sigma_b)$ & $2.92\,{\rm mm}\, \qty(9.74\,{\rm ps})$ & $3.3\,{\rm mm}\, \qty(10.9\,{\rm ps})$\\
        Filling pattern & -- & 1--400 buckets \\
        Electron lifetime & -- & $\sim10\,{\rm h}$ @ $160\,{\rm mA}$\\
        \end{tabular}
    \end{ruledtabular}
    \footnotetext[1]{Assuming the designed value}
    \footnotetext[2]{Converged by a consistent correction in operation} 
    \footnotetext[3]{$1.29\,{\rm nm}\,{\rm rad}$ assuming the designed energy spread of $0.0843\%$}
    \footnotetext[4]{Estimated from the obtained coupling constant and the model horizontal emittance}
    
\end{table*}

\begin{acknowledgments}
We thank all members of the Photon Science Innovation Center (PhoSIC) and QST NanoTerasu Center for their help and fruitful discussions over the whole commissioning period. 
We also greatly appreciate that many companies have been dedicated to accelerator complex constructions for several years.
We express our sincere thanks to Drs. K.~Soutome (JASRI, RIKEN) and T.~Fukui (RIKEN) for their helpful discussions.
We thank ThinkSCIENCE, Inc. (Tokyo, Japan) for English language editing.
\end{acknowledgments}

\appendix
\section{Magnet Alignment in the Storage Ring} \label{appendix:alignment}

The magnet alignment is vital to the linear optics correction in the storage ring. 
An alignment error causes additional dipole kicks in quadrupole magnets and quadrupole kicks in sextupole magnets.
Their contributions to COD and the beta function are:
\begin{equation}
    \Delta x^{\rm mon} = \frac{\sqrt{\beta_x^{\rm mon}}}{2 \sin{(\pi\nu)}} \cos{( \pi\nu - \Delta \psi_x )} \sqrt{\beta_x^{\rm mag}} k L \Delta x^{\rm mag},
\end{equation}
\begin{equation}   
    \Delta y^{\rm mon} = \frac{\sqrt{\beta_y^{\rm mon}}}{2 \sin{(\pi\nu)}} \cos{( \pi\nu - \Delta \psi_y )} \sqrt{\beta_y^{\rm mag}} k L \Delta y^{\rm mag},
\end{equation}
\begin{equation}
    \Delta \beta_x^{\rm mon} = \frac{{\beta_x^{\rm mon}}}{2\sin{(\pi\nu)}}\cos{2(\pi\nu - \Delta \psi_x )}\beta_x^{\rm mag} \lambda L \Delta x^{\rm mag},
\end{equation}
\begin{equation}
    \Delta \beta_y^{\rm mon} = \frac{{\beta_y^{\rm mon}}}{2\sin{(\pi\nu)}}\cos{2(\pi\nu - \Delta \psi_y )}\beta_y^{\rm mag} \lambda L \Delta x^{\rm mag},
\end{equation}
where 
$x^{\rm mon}$ ($y^{\rm mon}$) is the horizontal (vertical) COD at a BPM, 
$x^{\rm mag}$ ($y^{\rm mag}$) is the horizontal (vertical) COD at an alignment error magnet,
$\beta^{\rm mon}_x$ ($\beta^{\rm mon}_y$) is the horizontal (vertical) beta function at a BPM, 
$\beta_x^{\rm mag}$ ($\beta_y^{\rm mag}$) is the horizontal (vertical) beta function at an alignment error magnet,
$\Delta \psi_{x,y} \equiv \abs{\psi_{x,y}^{\rm mon} - \psi_{x,y}^{\rm mag}}$ is the difference in the betatron phase between the magnet and the BPM positions,
$\nu$ is the betatron tune,
$k$ is the quadrupole kick, $\lambda$ is the sextupole kick,
and $L$ is the magnet length.

The COD derived from the alignment errors should be converged within the range in which the steering magnets can apply counters.
Moreover, optics, such as the beta function, must also be converged within the linear responses.
The simulation considering the emittance, tune shift, COD, beta functions, and other optics parameters, showed that to substantiate the high-brilliance beam, the net magnet alignment $1\,\sigma$ error must be within $\pm 50\,\mu{\rm m}$: $\pm25\,\mu{\rm m}$ for quadrupole and sextupole magnets on the girder, $\pm50\,\mu{\rm m}$ for bending magnet on the girder, and $\pm 45\,\mu{\rm m}$ for between girders. 
To realize this, we performed three-step magnet alignments.

\begin{figure*}[htbp]
    \centering
    \includegraphics[width=0.8\linewidth]{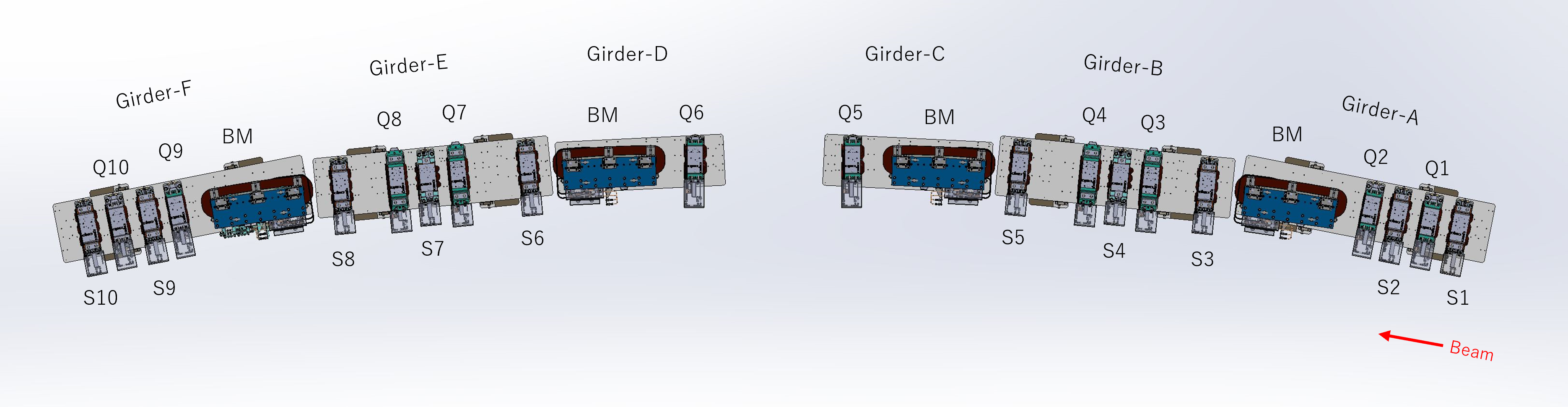}
    \caption{Magnet girder grouping. }
    \label{fig:girder}
\end{figure*}

First, the magnets were divided into six groups and aligned on six girders (Fig.~\ref{fig:girder}).
The magnetic centers for the quadrupole and sextupole magnets were aligned by the vibrating-wire-method (VWM)~\cite{10.1063/1.5086505} and fixed on each girder.
To monitor the center positions after the VWM alignment, the magnet positions were reflected into base points for laser tracker measurements.
The base points were mounted on top of each magnet and each girder.
After this step, the quadrupole and sextupole magnets were well-aligned on each girder, with an accuracy of $\pm 5\,\mu{\rm m}$ or better.
For the bending magnets, we measured the magnetic fields along the mechanical center line with a hole probe.
By evaluating the integrated dipole and quadrupole fields, the root-mean-square discrepancies from the designed values were found to be 0.09\% and 0.18\%, respectively.
Because of the B-Q combined type bending magnet, the dipole difference can be corrected by horizontal offsetting. 
We aligned the bending magnets on the girders by applying the offsets, and the maximum was smaller than $\pm$0.3\,mm.
The discrepancy of quadrupole components in the bending magnets was not mechanically corrected during this alignment phase.
The differences in dispersion and beta functions from the design as derived from the quadrupole component uncertainty were corrected by the linear optics correction using the beam responses as described in Sec.~\ref{subsec:loco}.

Second, in the girder alignment phase with the laser tracker, the girders were mounted on the storage ring floor, and the girder positions were fine tuned.
The floor is leveled with epoxy resin to ensure that the surface is flat and to mitigate the vibration transfer from the floor to the girders.
Referring to the laser-tracker base points on the floors and the walls of the storage ring tunnel, we evaluated the girder positions, ground levels, and rolls with the laser tracker.
The uncertainty of the laser tracker measurements is approximately $30\,\mu{\rm m}$ in our configuration, depending on the distances and angles between the magnets and the base points.
These works were performed from May to October 2022.

Third, in the confirmation phase in March 2023, we re-evaluated all magnet positions with the laser tracker after all the equipment was installed.

\begin{figure*}[hbt]
    \includegraphics[width=\linewidth]{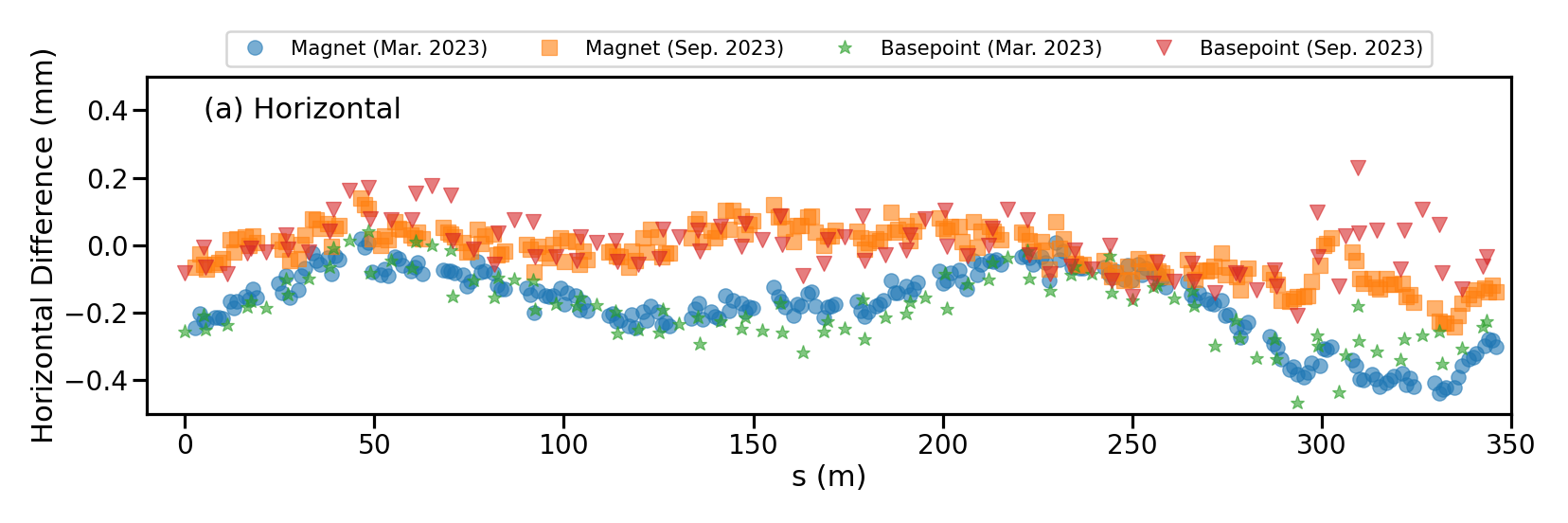}
    \includegraphics[width=\linewidth]{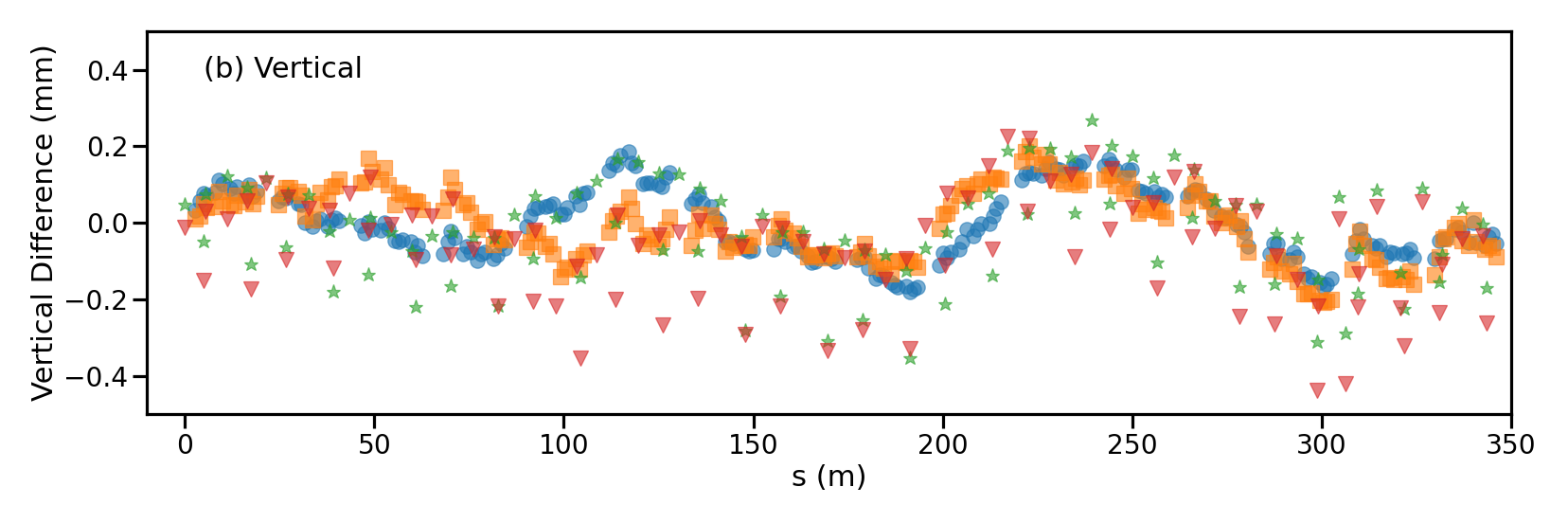}
    \caption{(a) Horizontal and (b) vertical displacements from the laser tracker measurement for March (blue and green) and September 2023 (red and orange). The vertical axis in the figure is the difference from the designed position, and the horizontal axis is the path length along the stored beam.}
    \label{fig:tracker_measurement}
\end{figure*}

Figure~\ref{fig:tracker_measurement}-(a) shows the horizontal displacement from the design, as measured by the laser tracker.
The blue circles show the magnet positions measured in March 2023, and sine-curve-like movement is visible.
The green stars show the base-point positions, and they also follow the sine-like curve.
However, the September 2023 data (orange circles and red stars) show that the displacements are within $\pm 0.2\,{\rm mm}$.
These variations can be explained by the seasonal deformation of the buildings and the grounds.
Because the girders are aligned with reference to the floor and wall base points, it is natural for the magnet position variation to obey the base-point seasonal variation.

In Fig.~\ref{fig:tracker_measurement}-(b), the vertical position deviation is stable between March and September 2023, in contrast to the horizontal position deviation.
Some base points on the floor are far from the designed positions, but they are also stable seasonally.
Because the stability of the buildings and ground influences the stable operation of the storage ring, we plan to continue to perform the laser tracker measurements regularly.

\begin{figure}[hbt]
    \centering
    \includegraphics[width=\linewidth]{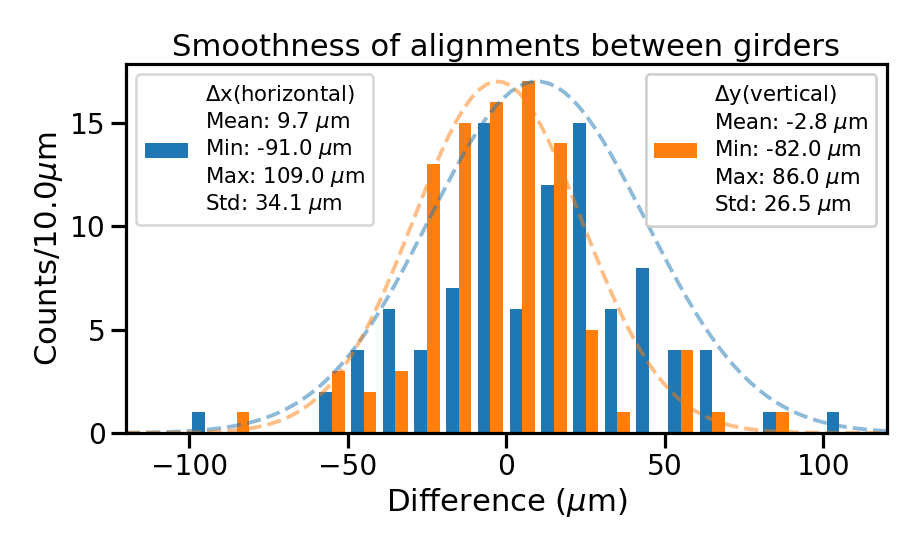}
    \caption{Relative displacements between the nearest girders from the laser tracker measurement on March 2023.
    The histograms of the horizontal and vertical alignment errors are distributed approximately centered. 
    }
    \label{fig:displacement_of_nextdoors}
\end{figure}

From the perspective of the stored electron beam, every magnet is on the trajectory of each girder, which is guaranteed by the VWM, even if the girder position is slightly far from the designed circular orbit compared with the buildings.
The alignment error problems with the additional kicks mainly arise from the step differences between the girders.
Hence, it is essential to prevent additional kicks so that the step-type alignment error is small.
Figure~\ref{fig:displacement_of_nextdoors} shows the relative magnet displacements between the neighborhood girders.
Each magnet girder is connected smoothly with $1\,\sigma$ within $34\,\mu{\rm m}$ ($26\,\mu{\rm m}$) in the horizontal (vertical), and this satisfies our requirement of being within $50\,\mu{\rm m}$.

\providecommand{\noopsort}[1]{}\providecommand{\singleletter}[1]{#1}%

\end{document}